\documentclass[a4paper,12pt]{article}
\usepackage{amssymb,amsmath,mathrsfs}
\usepackage[usenames,dvipsnames]{xcolor}
\usepackage{hyperref}
\usepackage{tensor}
\usepackage{graphicx}
\usepackage[left=1.2in,top=1in,right=1.2in,bottom=1in,headheight=0.8in,foot=0.5in]{geometry}
\usepackage[nodisplayskipstretch]{setspace} 
\usepackage[width=\textwidth]{caption}
\usepackage{pifont}
\newcommand{\cmark}{\ding{51}}%
\newcommand{\xmark}{\ding{55}}%
\usepackage[style=ext-authoryear-comp,maxcitenames=2,uniquename=false,backend=biber,uniquelist=false,articlein=false]{biblatex}
\DeclareNameAlias{sortname}{last-first}

\usepackage[indentafter]{titlesec}
\titleformat{name=\section}{}{\thetitle.}{0.8em}{\centering\scshape}
\titleformat{name=\subsection}[runin]{}{\thetitle.}{0.5em}{\bfseries}[.]
\titleformat{name=\subsubsection}[runin]{}{\thetitle.}{0.5em}{\itshape}[.]
\titleformat{name=\paragraph,numberless}[runin]{}{}{0em}{}[.]
\titlespacing{\paragraph}{0em}{0em}{0.5em}
\titleformat{name=\subparagraph,numberless}[runin]{}{}{0em}{}[.]
\titlespacing{\subparagraph}{0em}{0em}{0.5em}

\newcommand{\nocontentsline}[3]{}
\newcommand{\tocless}[2]{\bgroup\let\addcontentsline=\nocontentsline#1{#2}\egroup}
\hypersetup{
	colorlinks=true,         
	linkcolor=Black,          
	citecolor=MidnightBlue,
	urlcolor=MidnightBlue            
 }

\title{Explanatory Depth in Primordial Cosmology:\\ A Comparative Study of Inflationary and Bouncing Paradigms\vspace{-1.5em}\footnote{Forthcoming in \textit{The British Journal for the Philosophy of Science}}}
\date{}

\begin{document}
\maketitle
\setstretch{1.0} 

\begin{center}
\author{William J.~Wolf\footnote{Faculty of Philosophy, University of Oxford, UK. Email: william.wolf@philosophy.ox.ac.uk} and ~Karim P. Y. Th\'ebault\footnote{Department of Philosophy, University of Bristol, UK. Email: karim.thebault@bristol.ac.uk}}
\end{center}

\begin{abstract}
\noindent We develop and apply a multi-dimensional account of explanatory depth towards a comparative analysis of inflationary and bouncing paradigms in primordial cosmology. Our analysis builds on earlier work due to \textcite{Azhar:2021} that establishes \textit{initial conditions fine-tuning} as a dimension of explanatory depth relevant to debates in contemporary cosmology.  We propose \textit{dynamical fine-tuning} and \textit{autonomy} as two further dimensions of depth in the context of problems with instability and trans-Planckian modes that afflict bouncing and inflationary approaches respectively. In the context of the latter issue, we argue that the recently formulated trans-Planckian censorship conjecture leads to a trade-off for inflationary models between dynamical fine-tuning and autonomy. We conclude with the suggestion that explanatory preference with regard to the different dimensions of depth is best understood in terms of differing attitudes towards heuristics for future model building.
\end{abstract}
\tableofcontents
\setstretch{1.2}


\section{Introduction}

\noindent A heated debate has raged in contemporary cosmology regarding the scientific merits of the dominant inflationary paradigm. A small but influential minority of cosmologists have both questioned the justificatory basis for the predominate position of inflation and argued for an alternative paradigm based upon bouncing cosmological models. One axis of this debate relates to the comparison between the two approaches with regard to empirical support both in terms of \textit{prediction} and \textit{accommodation} of evidence. Most vividly, dispute along this axis has played out in exchanges between \textcite{Ijjas:2013vea,Ijjas:2014nta} and \textcite{Guth:2013sya,Guth:2017}. This aspect has recently received a detailed philosophical analysis from \textcite{Dawid:2021}.

Another axis of this debate concerns the relative \textit{explanatory} merits of the two approaches, taken both in comparison to each other and to the traditional hot big bang paradigm. In this context, it is worth noting that even the seemingly straightforward explanatory comparison between inflationary explanations and the hot big bang model proves controversial. Inflation was originally motivated by the observation that the hot big bang model involved \textit{implausible coincidences} which cried out for explanation \parencite{Smeenk:2005,azhar:2017}. However, explication of the basis for the superiority of the inflationary explanation in comparison to the `fine-tuned' hot big bang model is non-trivial. In particular, as persuasively argued by \textcite{Mccoy:2015}, a simple probabilistic framing of the explanatory virtues of inflation in comparison to the fine-tuned hot big bang model suffers from chronic ambiguities in defining probabilistic structure within a modern cosmological context. 

A first step towards a more satisfactory, non-probabilistic framing of the explanatory virtues of inflation over the hot big bang is to move away from a reliance on probabilistic notions and, following \textcite{Maudlin2007-MAUTMW}, focus on the fact that the inflationary explanation is a \textit{dynamical} one. In this spirit, \textcite{Azhar:2021} have recently argued that dynamical models with \textit{initial conditions fine-tuning} sacrifice \textit{explanatory depth} and that on this basis the explanatory superiority of inflation over the hot big bang can be established. 

In this paper, we will provide a \textit{multi-dimensional} analysis of the explanatory virtues of inflation in comparison to bouncing cosmologies. Our approach to explanatory depth in modern cosmology is founded on identifying  \textit{dynamical fine-tuning} and \textit{autonomy} as relevant dimensions of depth in addition to initial conditions fine-tuning. We take this to be a descriptively valid approach in this context since informal articulations of precisely these concepts are regularly invoked in the cosmology literature. In particular, with regard to dynamical fine-tuning and autonomy, we will demonstrate the particular relevance of these dimensions to the inflation vs.~bouncing cosmology comparison in the context of dynamical instabilities and the so-called trans-Planckian problem. 

Furthermore, we take there to be a good normative basis for treating depth along each of our three dimensions as an epistemic virtue since each of our three dimensions of explanatory depth are indicative of a form of \textit{explanatory modal robustness}. That is, in each of the dimensions a deeper explanation is such that, all else being equal, explanatory connections between explanans and explanandum will persist in a wider range of counterfactual scenarios, whether in terms of different initial conditions, different dynamical maps, or different realisations of phenomena at physical scales far away from the explanandum.

Our analysis will not lead us to a verdict with regard to the explanatory superiority of these two rival approaches to cosmology. Rather, we will seek to clarify the relevant terms of the debate, and in doing so, better understand the basis upon which scientists are in fact disagreeing. Furthermore, we will suggest that the different choices with regard to explanatory strategy have direct implications for the heuristics of model building in contemporary cosmology. The nature of the dispute can thus in part be understood in terms of a disagreement over different strategies regarding how best to constrain theoretical practice. Given the heavily unconstrained empirical environment of modern cosmology, such methodological diversity is well justified.        

\section{Primordial Paradigms: Bangs, Bounces, and Inflation}

\subsection{Hot Big Bang Model}

The Hot Big Bang (HBB) model is the standard paradigm for model building in cosmology.\footnote{For a more detailed overview of modern cosmology together with attendant philosophical issues see \textcite{Ellis:2014,Chamcham:2017,azhar:2017,Smeenk:2017}.} Its modern $\Lambda$CDM incarnation describes an expanding universe evolving according to the FLRW (Friedmann-Lema\^itre-Robertson-Walker) solution of General Relativity, that is composed of $\sim 5 \%$ baryonic matter, $\sim 26 \%$ non-baryonic cold dark matter, CDM, and $\sim 69 \%$ dark energy, $\Lambda$. This universe is both extraordinarily flat and homogeneous, with departures in homogeneity restricted to tiny density fluctuations of order $\sim 10^{-5}$ and precisely characterized by a nearly scale-invariant power spectrum.\footnote{These findings have been confirmed by main cosmology probes such as COBE, WMAP, and Planck \parencite{COBE, WMAP, Planck}, and received strong independent support from measurements of supernovae \parencite{Perlmutter}, baryonic acoustic oscillations (BAO) \parencite{aubourg2015cosmological}, galaxy rotation curves \parencite{Rubinreview}, gravitational lensing \parencite{Ellislensing}, Lyman-alpha forest \parencite{Weinbergforest}, galaxy clusters \parencite{Allenreview}.} 

The dynamics of any FLRW universe are generically captured by the two Freidmann equations:
\begin{equation}\label{velocity}
    H^2 \equiv \left(\frac{\dot{a}}{a}\right)^2 = \frac{1}{3}\rho - \frac{k^2}{a^2} + \frac{\Lambda}{3},
\end{equation}
\begin{equation}\label{acceleration}
    \dot{H}+ H^2 \equiv \frac{\ddot{a}}{a} = -\frac{1}{6}(\rho + 3p),
\end{equation}
where $a$ is the scale factor, $\rho$ is matter energy density, $p$ is pressure, $k$ is the spatial curvature, $\Lambda$ is the cosmological constant, and $H$ is the Hubble parameter. The HBB model takes the observed conditions and material constituents of the universe, and projects the evolution of the universe forward through these equations, as well as backwards towards an initial cosmic singularity. 

\subsection{Inflating and Bouncing Models}

Despite its successes, physicists believe that the HBB model ought to be modified and the most popular proposals for doing so fall into two main categories: inflationary models (the dominant paradigm) and bouncing models (a distant secondary option).\footnote{There are even more options than this, including string gas cosmology \parencite{Brandenberger:2008nx} and emergent universe models \parencite{Ellis:2002we}. However, we will not address these in this paper.} Such extensions primarily operate through the particular mass-energy content (i.e., the energy density and pressure) that these models place into the Friedmann equations, which then determines the subsequent evolution of the universe in terms of the velocity and acceleration of the scale factor. 

Inflation is a paradigm for building models within which the universe underwent a period of very rapid expansion at early times. There are, however, an extraordinary number of ways of implementing this paradigm. Physicists have cataloged and categorized \textit{at least} 74 different models of single-field inflation, not to mention more complicated multi-field models \parencite{martin2014encyclopaedia}. Here we will restrict our attention to single-field models as these represent the most common way of realizing the paradigm. Similarly, bouncing cosmology is not so much a single theory, but rather a paradigm of related models that implements the idea that the universe can transition from expansion to contraction and vice-versa from contraction to expansion. There are likewise many ways of modelling contracting and expanding scenarios. We will restrict our attention to models that pair ultra-slow contraction (`ekpyrotic contraction') with a non-singular bounce since these models are of current interest and distinguish themselves from inflation by avoiding singularities \parencite{ijjas2018bouncing}.\footnote{Alternative ways of constructing bouncing cosmologies include singular bouncing cosmologies and non-singular matter bounce cosmologies, amongst others \parencite{Brandenberger:2016vhg, Battefeld:2014uga}.} 

Inflation and bouncing models are primarily driven by a dynamical scalar field, $\phi$, with an associated potential, $V(\phi)$, which is coupled to gravity and dominates the mass-energy content of the universe during particular stages of evolution. For example, consider an action of the form: 
\begin{equation}
    S=\int \mathrm{d}^{4} x \sqrt{-g}\left[\frac{1}{2} R+\frac{1}{2} g^{\mu \nu} \partial_{\mu} \phi \partial_{\nu} \phi-V(\phi)\right].
\end{equation}
The dynamics of such models can be understood schematically by tracking the equation of state that results from the particular scalar fields and potentials that describe such scenarios. Generically, the equation of state for a scalar field is given by:
\begin{equation}
    w = \frac{p}{\rho} = \frac{\frac{1}{2}\dot{\phi}^2 - V(\phi)}{\frac{1}{2}\dot{\phi}^2+ V(\phi)},
\end{equation}
where $\dot{\phi}^2$ represents the kinetic energy of the scalar field. 

\subsubsection{Inflation Models}

\begin{enumerate}
    \item \textit{Inflation}: Inflation is driven by a scalar field (often called the `inflaton') with a \textit{positive} potential and these potentials are usually constructed so that there is a range of values for which the potential is relatively flat. When the potential is flat, the kinetic term $\dot{\phi}^2$ will be small as the field $\phi$ rolls down the potential function (`slow-roll inflation'). Under these circumstances, $V \gg \dot{\phi}^2$ and $w \approx -1$.
    \item \textit{Dynamics from inflation}: This corresponds to a period of accelerating expansion (i.e., $\ddot{a} \gg 0$ and $\dot{a} \gg 0$) as long as the potential dominates the equation of state. This leads to an exponential expansion of space where $a(t) \propto e^{Ht}$, which mimics the current epoch of dark energy dominated expansion \parencite{baumann2009tasi}. 
\end{enumerate}

\subsubsection{Bouncing Models}

\begin{enumerate}
    \item \textit{Bouncing models}: Bouncing models aim to construct a universe where expansion transitions to contraction, and vice versa. For instance, we are currently in a period of dark energy dominated expansion, which very well could be driven by a scalar field in a flat, positive range of its potential function (i.e., $w \approx -1$). Consider what happens if a relatively flat, positive potential $V$ evolves to become steep and \textit{negative}. Here, $\rho$ cannot be negative because it represents the energy density of the universe. The kinetic term $\dot{\phi}^2$ becomes large and is approximately equal to $V$, which leads to a small positive number in the denominator of $w$. Yet, for negative $V$, the pressure $p$ is also positive, which leads to a large positive number in the numerator of $w$. The result is that $w \gg 1$. 
    \item \textit{Dynamics from bouncing models}: When the scalar field begins rolling to negative values, the sign of the acceleration equation changes from $\ddot{a}>0$ to $\ddot{a}<0$ because $\frac{\ddot{a}}{a}=-\frac{8 \pi G}{3}\left(\dot{\phi}^{2}-V(\phi)\right)$, contrary to the inflationary case where $V$ is positive and dominates the equation. This decelerates the universe and eventually reverses expansion entirely. We can see this from recalling the first Friedmann equation $H^2 = \left(\frac{\dot{a}}{a}\right)^2=\frac{8 \pi G}{3}\left(\frac{1}{2} \dot{\phi}^{2}+V(\phi)\right)$ and recognizing that a large, negative $V$ offsets $\dot{\phi}$. Eventually, the negative acceleration will flip the sign of $H$ to negative, taking $\dot{a}>0$ to $\dot{a}<0$. Following this period of contraction, an appropriate modification of gravity (via the dynamics of the scalar field) leads to a non-singular bounce well before the Planck scale and the potential rolls back to a positive value, causing the universe to revert back to expansion \parencite{andrei2022end, ijjas2018bouncing, steinhardt2002cosmic, steinhardt2002cyclic}.
\end{enumerate}

\section{Explanatory Depth as a Theoretician's Virtue}

\noindent An adequate account of the dominant theoretical standpoint regarding model building in primordial cosmology requires consideration of the concept of scientific explanation. At first sight this observation may appear somewhat surprising since the HBB model provides a valid explanans for the relevant cosmological explananda under virtually any extant account of explanation \parencite{Earman:1999,Mccoy:2015}. If the strong theoretical preference for modifications of the HBB model is to be understood in explanatory terms, we are thus required to take a more nuanced approach to role of explanation in the context of primodial cosmology. The most obvious approach would be to understand the explanatory weakness of HBB model, and strength of its rivals, in probabilistic terms. The HBB model may explain flatness, but, by contrast, inflation both explains flatness \textit{and makes it overwhelmingly likely}. The problems with this style of argument have been carefully discussed in both the scientific and philosophical literature \parencite{Schiffrin:2012,Smeenk:2014,Curiel:2015,Mccoy:2015,Mccoy2017,Gryb:2021}. The key conclusion is that there exist chronic ambiguities in the non-arbitrary definition of probabilistic structure in a modern cosmological context. Depending on a judicious choice of formalisation one can justify both the statement that inflation is overwhelmingly likely and its negation. 

The failure of a probabilistic explanatory approach might plausibly be taken to point away from explanatory considerations altogether, or at the very least, downgrade their relevance \parencite{Mccoy:2015}. From an experimentally driven `empiricist standpoint', such a view appears to be well-justified. An alternative `theoretician's standpoint', by contrast, is that the inflationary explanation is indeed superior on account of having greater `depth'. The theoretician's search for an alternative explanation to that provided by the HBB model is then a search for \textit{deeper explanation}. How might we characterise the notion of explanatory depth more precisely? And what motivates theoretical cosmologists to be interested in deeper explanations?    

The first key observation, noted en passant by \textcite{Maudlin2007-MAUTMW}, is that what cosmologists are really looking for is a \textit{dynamical explanation}; an explanation that shifts the explanatory burden of the relevant observed facts from the initial conditions to the dynamical laws. Following \textcite[p.544]{baumann2009tasi}, we can accept that the big bang model is perfectly adequate if we assume initial conditions that are extraordinarily flat and homogeneous (with tiny inhomogeneities possessing just the right amplitude and features for structure formation), but ``a theory that explains these initial conditions dynamically seems very attractive". \textcite[p.116]{SteinhardtGuth} voiced a similar motivation in the initial development of inflation, noting that 
``from almost any set of initial conditions the universe evolves to precisely the situation that had to be postulated as the initial state in the standard model".

The aim of the deployment of the idea of explanatory depth in the context of modern cosmology is to make more precise the intuition behind such statements. In this context much valuable work has already been done in a recent paper by \textcite{Azhar:2021}, c.f.\ \textcite{Mccoy:2020}. Azhar and Loeb's account of explanatory depth focuses on explanatory generalisations that have a \textit{nomic character}, including dynamical explanations in our terms. The account is specifically fitted to the explanatory context of modern cosmology.\footnote{Azhar and Loeb explicitly acknowledge that their account extends earlier work on depth in explanatory generalisations due to \textcite{Hitchcock:2003}. They note, however, that the specific structure of the two schemes, and in particular the dimension of depth, are markedly different, although plausibly complimentary. There are also connections to the work of  \textcite{Weslake:2010} and \textcite{Ylikoski:2010}.} Under this account, one explanation is deeper than another when, for a fixed number of parameters and a single observable, there is a greater range of parameters that do not yield significant changes to the observable. 
For example, an equilibrium explanation provides a deep explanation in the sense of a lack of initial conditions fine-tuning, precisely because the explanandum of a final equilibrium state is suitably invariant under counterfactual values of the initial conditions part of the explanans. Stated another way, the less sensitive (and more robust) the explanatory relationship is to counterfactual values of parameters within the explanans, the deeper the explanation is. 

Azhar and Loeb provide quantitative means to make precise the highly intuitive idea that cosmological models with fine-tuned initial conditions, such as the HBB model, suffer from an explanatory deficiency. The explanations provided by HBB models are such that parameters within the explanans can only take on a relatively small range of values without dissolving the explanatory relationship with the relevant explanandum. By contrast, the extensions of the HBB model are such that there is a greater range of parameters under which the relevant explanatory connections are sustained.  Azhar and Loeb persuasively argue both towards the general claim that finely-tuned explanations sacrifice explanatory depth and the specific claim that the key reason for cosmologists preferring inflation to the HBB can be understood in these terms. 
We are in agreement with this analysis and will provide an overview of the relevant physical ideas along similar lines in Section \ref{ICandED} (we refer the reader to the original paper for the quantitative framing in the inflation case). 

At a more abstract level it will prove worthwhile to note that as well as evidently being a \textit{descriptively adequate} account of the explanatory situation in cosmology, the Azhar and Loeb analysis of initial conditions fine-tuning and explanatory depth offers a clear \textit{normative basis} for depth qua lack of initial conditions fine-tuning to be treated as an \textit{epistemic virtue}. This is because, by definition, an explanation with less initial conditions fine-tuning will be one in which, all else being equal, the explanatory connection between explanans and explanandum will persist in a range of counterfactual scenarios with different initial conditions. One can thus understand the epistemic virtue of an explanation that is deep in the sense of avoiding initial conditions fine-tuning in terms of its \textit{modal robustness}. That is, there are a greater variety of ways the world could have been in which the relevant explanatory connection would still obtain.  We will return to the generalisation of this idea that the modal robustness of an explanation is an epistemic virtue in what follows.


There are four aspects which we take need to be added to the account to allow for a comparative study of inflation and bouncing paradigms in terms of explanatory depth. The first is an advocation of a multi-dimensional account of explanatory depth. Following the discussion of \textcite{Ylikoski:2010} and \textcite{Weslake:2010}, c.f.\ \textcite{Jackson:1992}, we take explanatory depth to be a non-unitary concept with different dimensions relevant to different domains. We further take cosmology to be a domain in which there are at least three relevant dimensions. 

The second aspect of our proposal is the suggestion that the absence of \textit{dynamical} fine-tuning is a relevant dimension of depth in the cosmological context. This involves a moderate modification of Azhar and Loeb's notion of depth by shifting focus from fine-tuning of the \textit{initial conditions} to fine-tuning of the \textit{dynamical maps}. Whereas the former are understood as points in state space that specify the state of the system at some initial time, the latter are one parameter groups of transformations that map the state of the system at one time onto the state of the system at some other time.\footnote{The concept of a dynamical map is drawn from the widely applicable framework of Dynamical Systems \parencite{Broer:2011}.}
In analogy with the previous definition of initial conditions fine-tuning, dynamical fine-tuning is defined such that one explanation is deeper than another when, for a fixed number of parameters and a single observable, there is a greater range of appropriately similar dynamical maps that do not yield significant changes in the observable.  For example, a renormalization group explanation provides a deep explanation in the sense of a lack of dynamical fine-tuning, precisely because the explanandum of critical behaviour is suitably invariant under a variety of counterfactual forms of the fundamental Hamiltonian that enters into the explanans.\footnote{This is a greatly simplified illustrative example. For more on renormalization group explanations see  \textcite{batterman:2000,batterman:2002,Reutlinger2014-REUWIT,franklin:2018}.} In what follows we will argue that the key explanatory weakness of bouncing models is lack of depth along the dimension of dynamical fine-tuning. 

The articulation of the normative basis for depth qua lack of dynamical fine-tuning to be treated as an epistemic virtue parallels that we just gave for initial conditions fine-tuning. An explanation with less dynamical fine-tuning will be one in which, all else being equal, the explanatory connection between explanans and explanandum will persist in a range of counterfactual scenarios with different dynamical maps. Thus, once more, one may understand the epistemic virtue of an explanation that is deep in the relevant sense of lacking dynamical fine-tuning in terms of its modal robustness.

In correspondence to our discussion regarding probabilistic structure above, we take there to be good reasons to avoid seeking to fully formalise the sense of modal robustness that occurs in both initial conditions and dynamical fine-tuning. Such formalisation would require a precise formulation of both the relevant space and a measure on that space. For initial conditions it is plausible to identify the relevant space in terms of initial conditions for the perturbed Friedmann equations. However, there are serious ambiguities in the definition of the measure on such a space -- on this see in particular \textcite{Schiffrin:2012} and \textcite{Curiel:2015}. In the context of dynamical maps, while in principle it might be possible to conceive of a space of dynamical maps, any concrete formulation would require strong theoretical constraints on model building. Such a formalisation would be highly vulnerable to volatility in theoretical understanding, notwithstanding the (presumably intractable) question of finding and justifying a measure.\footnote{Thanks to an anonymous referee for helpful comments on this point.}

The third aspect of our proposal involves characterisation of a further dimension of depth, this time a more significant departure from Azhar and Loeb's account. Here we are drawing partial inspiration from the idea of \textcite{Weslake:2010} that \textit{autonomy} is a significant dimension of explanatory depth. In our characterisation, one dynamical explanation is deeper than another when the explanatory connection between the explanans and explanandum is insensitive to the breakdown of our dynamical modelling frameworks or laws in regimes at very different scales from that of the explanandum.\footnote{Scale here might refer to spatial scale, energy scale, temporal scale, or numerical competent scale. Autonomy is in our sense is closely connected to the conception of scale-relativity discussed by \textcite{Ladyman:2020} and \textcite{Cretu:2022}.} Paradigmatically, an explanation will be deep in the sense of autonomy when the scale of the explanans and explanandum are closely matched and an explanation will be shallow in the sense of autonomy when the explanans make reference to a much smaller spatial scale than the explanandum, and the relevant explanatory connection is highly sensitive to assumptions regarding this scale.\footnote{For Westlake autonomy is the view that it is possible for non-fundamental sciences to provide deeper explanations than fundamental science on the basis of a greater degree of abstraction, where abstraction is applicability to a greater range of types of physical systems. Our conceptualisation of autonomy, by contrast, is inspired by  the idea of applicability to a greater range of realisations of the mico-physical structure underlying one type of physical system. The distinction is thus broadly equivalent to that between universality and robustness   \parencite{batterman:2000,batterman:2002,gryb:2020,Palacios:2022}.} 

The most vivid illustration of dynamical explanations that are deeper in the sense of autonomy are the explanations provided in continuum fluid mechanics, c.f.\ \textcite{Darrigol:2013,Batterman:2018}. Such explanations are successful precisely because their objects are macro-scale explananda for which they provide macro-scale explanans. As such, they avoid reference to many-body molecular scales which are outside the validity of the relevant continuous fluid dynamical laws (since they are at a smaller spatial scale) and outside the tractability regime of the more fundamental molecular dynamics (since fluids have numbers of components of the order of the Avogadro number). 

The articulation of the normative basis for depth qua autonomy to be treated as an epistemic virtue again draws upon the idea of modal robustness. An explanation which is more autonomous will be one in which, all else being equal, the explanatory connection between explanans and explanandum will persist in a range of counterfactual scenarios where the realisation of phenomena at physical scales far away from the explanandum is very different. This idea can be clearly illustrated by the example of continuum fluid mechanics where the explanations provided are autonomous precisely because they are insensitive to a huge variety of possible micro-physical realisations of the underlying molecular-hydrodynamics.   

It is important to note that our conception of autonomy as a dimension of explanatory depth does not make reductive explanations \textit{necessarily} shallow in the sense of autonomy. Rather, our approach cautions against a form of reductive explanation that proceeds by connecting explananda at one scale, to explanans that require highly specific approaches to modelling at very different scales. In general, we have good reasons to expect such explanations to be less modally robust precisely because they require specific modelling assumptions to hold across multiple scales. It is certainly possible in principle for explanations to be both deep in the sense of autonomy and reductive in the sense of having the explananda and explanans defined at different scales.\footnote{In particular, there can be circumstances where the relevant cross-scale explanatory connection does not display a high level sensitivity to model assumptions regarding the scale of the explanans. We will consider an example of such an explanation in the context of explanations of Hawking radiation based upon modified dispersion relations in Section \ref{InflatingPlanck}. Arguably, this is also precisely what reductive accounts of renormalization group explanations aim to demonstrate \parencite{reutlinger:2017,saatsi:2018,franklin:2018,castellani:2022}.} Moreover, there are situations in science where it turns out to be unavoidable to connect phenomena at one scale to models and theories at a very different scale, even when the theory at that scale is not well established, and the robustness of the explanatory connection is not yet understood. A particularly good historical example of such a situation is the requirement for early twentieth century physicists to move beyond classical physics into the then highly speculative domain of quantum physics in order to account for the thermal properties of solids \parencite{Darrigol:2009}. We take there to be good reasons to expect that, all else being equal, autonomous explanations will more reliable. This does not, however, rule out there being situations where all else is not equal, and only a non-autonomous explanation is in fact available.\footnote{We are very appreciative of comments from an anonymous referee that have prompted us to clarify our thoughts regarding the potential for failure of autonomy in scientific explanations. Ultimately, we take there to be good general formal and physical reasons to expect autonomous explanations to, in principle, be possible in physical contexts in which Effective Field Theory (EFT) can be applied. We will return to this point, and its relation to the potential breakdown on the EFT framework in primordial cosmology, in Section \ref{InflatingPlanck}.}

In the cosmological context, the explananda are tied to the scale of the CMB and the relevant dynamical modelling framework is the perturbed Friedmann equations together with quantum field theory. An explanation that is deep in the sense of autonomy will, in this context, be one in which the explanatory connection between the relevant explanans and explanandum does not require highly specific approaches to modelling scales far away from the CMB scale. In what follows we will argue that inflationary models generically lose out to bouncing models in the autonomy dimension of explanatory depth due to the so-called trans-Planckian problem, which requires reference to scales far beyond the the CMB scale. Interestingly, we will also argue that the subset of inflationary models that are autonomous, those that satisfy the so-called trans-Planckian censorship conjecture (TCC), are found to sacrifice explanatory depth along the dimension of dynamical fine-tuning. Autonomy and dynamical fine-tuning, as dimensions of explanatory depth, are thus precisely those required to elucidate the structure of the relevant debates since the different putative cosmological explanations prove to excel along the different dimensions defined within our account of explanatory depth.  

This brings us to the fourth aspect of our account, which is drawn from the problem of comparing rival explanations that excel along different dimensions of explanatory depth. On our view, the choice for theoretical cosmologists implied by such comparisons is best understood in terms of differing attitudes with regard to heuristics for future model building. That is, a  choice between explanations with different forms of \textit{heuristic fecundity}. Most straightforwardly, the reason why explanations that lack depth qua initial conditions fine-tuning are so unsatisfactory is, at least in part, due to the heuristic sterility of explanations of phenomena that appeal to special initial conditions. A more complex choice, that we will argue to be relevant to our particular context, is between explanations that are deeper along the dimensions of autonomy and dynamical fine-tuning. The heuristic value of an autonomous but dynamically fine-tuned explanation can be understood in terms of the positive heuristics provided for theoretical model building in a constrained space within limitations on both the realm of relevant empirical phenomena and the possible dynamical structures that can be implements. By contrast, the value of an explanatory approach that is deep in virtue of not being dynamically fine-tuned, but shallower in virtue of lack of autonomy, might be understood in broadly empiricist terms. We will return to these issues in the final section. 

In conclusion, our approach to explanatory depth in modern cosmology is founded on identifying initial conditions fine-tuning, dynamical fine-tuning, and autonomy as the relevant dimensions of depth. We take this to be a descriptively valid approach in this context since informal articulations of precisely these concepts are regularly invoked in the cosmology literature. Furthermore, we take there to be a good normative basis for treating depth along each of these dimensions as an epistemic virtue since each of our three dimensions of explanatory depth are indicative of a form of \textit{explanatory modal robustness}. That is, in each of the dimensions a deeper explanation is such that, all else being equal, explanatory connections between explanans and explanandum will persist in a wider range of counterfactual scenarios, be these in terms of different initial conditions, different dynamical maps, or different realisations of phenomena at physical scales far away from the explanandum. 
Finally, while we do not take there to be epistemic norms that dictate the choice between explanations that excel along different dimensions of depth, we think that there is a good basis to draw connections between such choices and differing pragmatic attitudes of theorists with regard to heuristics for future model building.

\section{Initial Conditions and Explanatory Depth}
\label{ICandED}
\subsection{Flatness Problem}

The \textit{flatness problem} can be characterized as the realization that, for the universe to possess its observed flatness today, the initial value of the density $\rho$ had to be extraordinarily close to what is known as its critical value \parencite{Holman:2018}. The critical value for density $\rho_c$ is the unique value for which $k=0$ according to the first Friedmann equation and can be written as the ratio $\Omega = \rho/\rho_c$. This allows one to write the so-called curvature parameter $\Omega_k$ in the following suggestive way:
\begin{equation}
    \Omega_k = 1 - \Omega = \frac{k}{a^2H^2}
\end{equation}
The scale factor, $a$, is proportional to the actual size of the universe and the Hubble parameter, $H$, is inversely proportional to how far an observer can actually see (i.e., in units where $c=1$, the Hubble horizon is simply $H^{-1}$). Thus, this parameter can be roughly understood as being proportional to the ratio of the apparent size of the universe to its actual size. If $\Omega = 1$, $\Omega_k = 0$ and $k=0$, and the universe is flat. However, this is an unstable fixed point for energy inputs like radiation or matter in the HBB model. 
Thus, the curvature parameter $\Omega_k$ will diverge over time \textit{regardless} of what its initial value was as $a$ and $H$ evolve in the HBB model. A universe that is observed to be essentially spatially flat today requires a density that was extraordinarily close to the critical value at earlier times (within $10^{-55}$ of the critical value when extrapolating back to GUT scale energies \parencite{baumann2009tasi}). 

The preceding discussion on inflationary dynamics immediately equips us to see how an inflationary epoch will offer a significant increase in explanatory depth when compared to the HBB model. Recall that $H = \dot{a}/a$ and that during inflation, $a(t) \propto e^{Ht}$. Thus, $H$ is approximately constant while $a$ grows exponentially, and consequently, $\Omega_k$ is driven towards zero. This is analogous to how curved spaces can appear flat when the space we are considering is sufficiently small compared to the actual radius of curvature. Inflationary dynamics completely turn the tables. Rather than requiring a finely-tuned, specially chosen density value to explain why the universe we see today is flat, seemingly any initial density value will correspond to a flat universe. The HBB drives the universe away from flatness while inflation drives the universe towards flatness. 
When compared to the HBB model, inflation offers a deeper explanation because it allows for significantly more variation in the values of the initial conditions without dissolving the explanatory relationship with the explanandum. 


Does this analysis carry over to bouncing models? At first glance, it might seem surprising that it does. After all, a contracting universe is seemingly the opposite of an expanding universe. If exponential expansion flattens the universe, how could slow contraction accomplish the same? Furthermore, as $a$ gets smaller during contraction shouldn't that amplify the curvature parameter rather than suppress it? The key insight is that during a slow contraction the behavior of $H$ changes as well. A simple manipulation of the Friedmann equations shows that, when the equation of state $w \neq -1$, $H^{-1} \propto a^\epsilon$, where $\epsilon = \frac{3}{2}(1 + w)$. For a contracting universe with an equation of state $w>>1$, this means that the curvature parameter can be written as $\Omega_k \propto a^{2\epsilon}/a^2$ \parencite{ijjas2018bouncing}. While $a^2$ decreasing in the denominator would seemingly blow up the curvature parameter, $\Omega_k$ is actually suppressed because the numerator $a^{2\epsilon}$ decreases faster. That is, the universe visible to an observer shrinks faster than the universe itself. Thus, a period of slow contraction mimics the effects of an inflationary period and likewise drives the universe towards flatness. Furthermore, this mechanism's effectiveness has been confirmed with detailed numerical simulations \parencite{ijjas2020robustness}. Bouncing models which utilize this contraction mechanism offer an account of explanatory depth that is similarly compelling to their inflationary competitors. 

\subsection{Horizon Problem}

The universe is also remarkably homogeneous, with departures from homogeneity showing up only at the level of $\sim 10^{-5}$. Homogeneity by itself is not a problem. It would not be surprising for a system of particles in causal contact to attain conditions and properties that are nearly homogeneous, but the fact that the vast majority of the universe exists in causally disconnected patches makes observing such homogeneity a genuinely striking puzzle. This is known as the \textit{horizon problem}. The Hubble horizon forms a causal past light cone for each observer. Many of the points in the universe that we see today lie outside outside each other's present Hubble horizons, while displaying the same remarkable homogeneity. It has been estimated that at the time of recombination when the CMB photons first started streaming, the universe consisted of $\sim 10^{83}$ causally disconnected regions \parencite{Mukhanov:2005sc}. 

Inflating and bouncing cosmologies approach this problem similarly. Rather than simply asserting that the initial conditions of the universe were such that causally disconnected regions happen to be homogeneous, they provide a mechanism such that these regions of the universe share a causal past, before then explaining why these regions $\textit{appear}$ to be causally disconnected now. 

Prior to an inflationary phase, distant points in the universe share a causal past. During inflation, the rapid exponential expansion of space shrinks the so-called co-moving Hubble horizon $(aH)^{-1}$. Intuitively the co-moving Hubble horizon represents the fraction of the universe that is observable. Following exponential expansion, this co-moving Hubble horizon shrinks significantly, meaning that regions that were in prior causal contact now appear to be outside each other's past light cones. Inflation needs approximately 60 $e$-folds (i.e., the time interval in which the exponentially growing scale factor grows by a factor of $e$) to solve the horizon problem \parencite{baumann2009tasi}.

Coming to bouncing models, we find again that inducing a period of slow contraction produces similar behavior. Slow contraction causes the co-moving Hubble radius to shrink. Rather than shrinking via the exponential expansion of $a$ as with inflation, the co-moving Hubble radius shrinks because $a^\epsilon$ declines faster than $a$. This is again due to the very different behavior of $H$ given matter-energy content with an equation of state $w >> 1$. Upon transitioning to a subsequent expanding phase, the co-moving Hubble horizon proceeds to grow again, which results in regions that were previously in causal contact re-entering the horizon, while appearing as if they were never in causal contact \parencite{ijjas2018bouncing}.



Both of these accounts allow the universe we observe today to share a common causal past, yet as \textcite[p.10]{ijjas2018bouncing} explain, ``removing the causal impediment is necessary but not sufficient to explain why the energy
density distribution was so smooth at the time of last scattering". The aforementioned dynamical mechanisms responsible for resolving the horizon problem manage to explain this as well. Rather than postulating what would need to be very finely-tuned initial conditions within causally disconnected regions of the universe, both paradigms allow for a far larger variance in initial parameter values that eventually lead to the same observable state of interest. 
Even if there were somewhat significant inhomogeneities in these causally connected regions, the dynamics of inflation models inflate them away \parencite{Brandenberger:2016uzh, East:2015ggf}, whereas conversely the dynamics of bouncing models dramatically shrink them \parencite{ijjas2018bouncing}, with the end result being that the universe is homogenized.\footnote{Of course, these cosmological models are not compatible with \textit{any} conceivable initial conditions. However, in both cases, the acceptable range of initial conditions that produce the universe we observe is dramatically enlarged by orders of magnitude.} Here again, dynamics drive the explanatory power and depth of the respective models, and there is a significant reduction in the fine-tuning of initial conditions needed to account for the observed state of the universe.

\subsection{Scale-Invariant Density Perturbations}

The inhomogeneities in the CMB are extremely important to cosmology because it is these density perturbations that ultimately seed the large-scale structure in the universe. It had long been argued that primordial density perturbations should be scale-invariant ($n_s = 1$) \parencite{Harrison, PeeblesYu, Zeldovich}.\footnote{This property has been measured and is frequently discussed as the scalar-spectral index $n_s$. Planck has measured this to be $n_s =  0.9649 \pm 0.0042$, with perfect scale-invariance corresponding to $n_s = 1$ \parencite{Planck}.} 
However, within the HBB model it is not at all clear where these density perturbations come from. Of course, you can put finely-tuned inhomogeneities in the initial conditions, but this would add yet another implausible degree of fine-tuning.


The prediction of a nearly scale-invariant spectrum of density perturbations is counted as one of the most important successes of the inflationary paradigm. Within a few years of the theory first appearing, it became clear that inflation could source these perturbations and several researchers had independently derived a nearly scale-invariant spectrum of fluctuations \parencite{Mukhanov:1981xt, Press_1980, Guth1982, HAWKING1982295, Bardeen1983}. Intuitively, these perturbations represent tiny quantum mechanical variations in the field values of the inflaton itself. We can use standard quantum field theory to quantize these perturbations and compute their quantum statistics, resulting in the power spectrum \parencite{baumann2009tasi}:
\begin{equation}
    \Delta_{\mathcal{R}}^{2}(k)=A_{\mathrm{s}}\left(k / k_{*}\right)^{n_{\mathrm{s}}-1},
\end{equation}
where $\Delta_{\mathcal{R}}^{2}$ is the power spectrum of density fluctuations $\mathcal{R}$, $A_s$ is the amplitude, $k$ is the fluctuation mode, $k_{*}$ is a reference length scale usually taken to be the horizon crossing, and $n_s$ is the scalar spectral index. ${n_{\mathrm{s}}}$ can be computed for any inflation model from analyzing its dynamics and $n_s \simeq 1$ is a generic feature of single-field inflation models that satisfy the typical constraints in the inflationary paradigm (such as having a relatively flat potential with a valid slow-roll approximation). 

Bouncing paradigms predict a nearly scale-invariant spectrum as well through a similar procedure (with some important differences, as we shall see later). Similar to the flatness and horizon problems, both inflating and bouncing paradigms invoke dynamics that drive the values of important features of the universe towards those that are actually observed, and both offer similar increases in explanatory depth over the standard HBB model by shifting the burden of explanation from finely-tuned initial conditions to dynamics.

\section{Dynamics and Explanatory Depth}


\noindent Dynamics, like initial conditions, can be varied within the explanans. This may seem strange at first since we are used to working with fixed dynamical laws while varying parameters such as initial conditions. However, inflation and bouncing cosmologies are not themselves specific theories with fixed dynamical laws, but rather paradigms with many different dynamical realizations. In other words, we are interested in how effective these research programmes are at providing a greater range of dynamical maps appropriately suited to describing the observable universe, for a given set of parameters and observables. In this context, we can examine the space of dynamical realizations within the paradigms and evaluate these competing paradigms on the sensitivity of their explanatory relationships to different dynamical implements. We will show that this is one manner in which these paradigms start to diverge in their explanatory depth, with inflation emerging as the deeper explanation in terms of this dimension of dynamical fine-tuning. Furthermore, as we shall see, cosmologists who favour inflation have long pointed to this as a significant virtue of the inflation paradigm.

\subsection{Primordial Gravitational Waves}

A positive detection of primordial gravitational waves, or tensor perturbations is a major goal in observational cosmology. Two of the primary reasons these perturbations are so significant are as follows: (i) they produce a distinctive B-mode polarization that cannot be mimicked by the types of scalar perturbations we have already detected \parencite{Zaldarriaga:1996xe} and (ii) such a distinctive signature would be seen by many cosmologists as strong evidence for inflation because inflation generally predicts significant production of primordial gravitational waves \parencite{Baumann:2009mq}. It is important to note that such a detection does not uniquely single out inflation. As \textcite{Brandenberger:2011eq} emphasizes, primordial gravitational waves can be produced both by topological defects in standard HBB cosmology or in particular realizations of other early universe paradigms. Thus, if tensor perturbations were detected, one would have to carefully examine additional data points such as the tensor spectral tilt to differentiate competing theories. 

Such tensor perturbations can be directly related to the energy scale of an inflating or contracting mechanism because the ratio between tensor and scalar perturbations $r$ can be manipulated to directly constrain $V$ and the energy scale of such a mechanism \parencite{baumann2009tasi}. Inflation is generally expected to occur at near GUT-scale energies, leading to a relatively high production of tensor perturbations and tensor-scalar ratio $r$. On the other hand, the slow contraction mechanism employed by the kinds of ekpyrotic bouncing models we are considering\footnote{It should be noted that some other kinds of bouncing models that, such as a pure matter bounce scenario, can lead to significant production of primordial gravitational waves \parencite{Brandenberger:2011eq}.} occurs at lower energies that are much further away from the Planck scale, leading to significantly lower expectations for tensor perturbations, which would make $r$ unobservably small \parencite{Ijjas:2019pyf}.

The most recent Planck constraints indicate that $ r < .10 $ \parencite{Planck}. Clearly, this is not a problem for bouncing models. However, these recent constraints actually rule out many of the simplest and most studied inflation models that broadly fall under the a category known as `power-law inflation'. Following the Planck results, detailed assessments of the paradigm have found that `plateau inflation' models are now strongly favoured by the data \parencite{Chowdhury:2019otk, Martin:2015dha, Planck:2018jri}. While these models are not prima facie unreasonable\footnote{Although, they have been criticised for requiring more parameters and fine-tuning than power-law models in order to achieve the same desired outcomes \textcite{Ijjas:2013vea}.}, this episode illustrates that the inflation paradigm has had to invoke some non-trivial degree of dynamical fine-tuning to account for present observational constraints on the tensor-scalar ratio. 

\subsection{Scale-Invariant Density Perturbations}

While generically predicting unobservable tensor perturbations and a low $r$ value is certainly a point for the bouncing paradigm, things get a little more complicated when coming back to the scalar perturbations. While both inflation and bouncing paradigms can produce results consistent with the scale-invariant spectrum of density perturbations seen in cosmological probes, inflation does so in a more natural way.

In an expanding, inflating universe the growing scalar modes that are understood to be the all-important seeds of structure formation are actually decaying modes in the corresponding time-reversed, contracting universe. Similarly, the growing modes in a contracting universe map onto the decaying modes in the corresponding expanding universe \parencite{Lehners:2007ac, Creminelli:2004jg}. The dynamics responsible for inflation naturally source scale-invariant density perturbations through the typical growing modes. However, a bouncing cosmology needs to reckon with the fact that the growing modes during the contraction phase become decaying modes during subsequent expansion, but the decaying modes that would naturally grow in the subsequent expansion have already decayed away. 

One way to solve this problem is to introduce an additional, `spectator' scalar field that couples to the ekpyrotic scalar field \parencite{Lehners:2007ac, Levy:2015awa}. While the details are beyond the scope of this paper, the coupling of the spectator and ekpyrotic fields can generate a scale-invariant spectrum of density fluctuations. Other solutions include choosing particular matching conditions to match growing modes in the contracting phase to growing modes in the expanding phase, but this is arguably less desirable as it requires very specific choices of matching conditions \parencite{Brandenberger:2016vhg}. 

There is thus is a sense in which the bouncing cosmology paradigm requires dynamical fine-tuning in a way that the inflation paradigm does not. Inflation and its many dynamical realizations generically predict density perturbations with the features we observe, whereas the basic dynamical realizations of bouncing cosmologies require supplementation in the form of additional dynamical variables to produce the same results.

\subsection{Avoiding Instabilities}

Perhaps the biggest hurdle that bouncing models have to overcome is the existence of instabilities. Within physics, `instability' can have a few different meanings. It could refer to an unstable fixed point, such as we saw in the example of the flatness problem. This is not in and of itself disqualifying as it just means that we don't expect the system to remain in its state for very long. Instabilities can manifest in far more concerning ways though, in the form of an unbounded Hamiltonian. These instabilities are frequently called `ghost' or `gradient' instabilities and are considered to be so problematic because they are both perturbatively ill-defined and can lead to the infinite production of non-physical, negative energy states \parencite{Rubakov:2014jja, Wolf:2019hzy}. These frequently manifest themselves in the form of `wrong-signed' terms in a theory's Lagrangian, such as a minus sign in front of the kinetic term. Theories with such instabilities are not generally considered to be physically viable. 

This pathological behavior can be traced to the fact that bouncing cosmologies violate the \textit{null energy condition} (NEC). The NEC holds that for any form of material content, $p+\rho \geq 0$. Inflation does not violate this constraint as this condition holds during an expansion phase with $w \approx -1$; however, bouncing cosmologies necessarily violate this condition when they transition from contraction to expansion. That is, $\dot{H} \propto -(p+\rho) \leq 0$ during contraction, and flipping $\dot{H}$ from $\dot{H} < 0$ to $\dot{H} > 0$ when contraction reverses to expansion requires violating this energy condition \parencite{ijjas2018bouncing}.

This problem is typically approached by introducing non-standard kinetic terms into the relevant scalar fields and/or introducing modifications to gravity that become relevant during the bounce phase \parencite{Ijjas:2016vtq, Ijjas:2016tpn, Easson:2011zy, Cai:2012va}.\footnote{This problem of constructing stable solutions in bouncing cosmologies is subtle. See also \textcite{Cai:2016thi, Cai:2017dyi, Cai:2017tku, Libanov:2016kfc, Kobayashi:2016xpl, Creminelli:2016zwa} for various approaches and developments in solving this problem.} This usually takes inspiration from Horndeski gravity, which is the most general form of a scalar-tensor theory of gravity leading to second order equations of motion \parencite{Horndeski:1974wa}. In particular, one way of doing this makes use of the so-called $\mathcal{L}_4$ interaction, which includes a non-minimal coupling between the scalar field and the Ricci scalar as well as non-standard kinetic terms. This particular variant of modified gravity within non-singular bouncing models allows for a stable violation of the NEC before, during, and after the bounce phase, free of pathologies.

This reflects an interesting way in which the dynamical realizations of the bouncing paradigm need to be dynamically fine-tuned (i.e., introduce highly specific modified gravity dynamics). Not all dynamical fine-tuning is bad. When examining the space of dynamical realizations of these paradigms, it is not at all problematic for \textit{empirically motivated} dynamical fine-tuning to enter the picture as new observations further constrain models. Indeed, this can actually be desirable as it narrows the space of acceptable theories and helps theorists and experimentalists focus on those models which are more likely to be successful. However, dynamical realizations of the bouncing paradigm need not only be fine-tuned to accord with some observations, as does inflation, they must also be dynamically fine-tuned to be viable in principle. 

Inflation possesses more explanatory depth in the dimension of dynamical fine-tuning because the paradigm itself can sustain the explanatory relationship with the observable universe in a way that is far less sensitive to specific choices with regard to the structure that dictates the relevant dynamical maps. This relative insensitivity to dynamical structure was in fact understood quite early on in the theory's development. In particular, \textcite[p.180]{Linde:1983gd} argued in a seminal paper that our conclusions regarding inflation's ability to produce a universe like the one we observe is ``almost model-independent" and that ``inflation occurs for all reasonable potentials V($\phi$)", with the only real conditions being that the potential must be reasonably flat and operate at sufficient energy scales. This reasoning was also echoed in \textcite[p.112]{Guth:2013sya} in their response to \textcite{Ijjas:2013vea} as they noted that inflation generates ``generic predictions" that correspond to our observed universe and that these generic predictions ``are consequences of simple inflationary models" and depend only on ``the energy scale of the final stage of inflation".  Bouncing models and their relevant dynamical structures, on the other hand, necessarily need to be wedded to very specific modified gravity dynamics, along with any associated baggage, to maintain their physical and explanatory viability. Indeed, leading proponents of inflation have critically noted that that the kinds of modifications utilized in these efforts seem to be inordinately complicated and fraught with difficulties \parencite{Linde:2014nna}. The bouncing paradigm thus requires more dynamical fine-tuning than the inflationary paradigm and this reflects the reality that it is simply a much more difficult and non-trivial task to construct physically viable models within the bouncing paradigm.

\section{Autonomy and Explanatory Depth}

\noindent Recall that one dynamical explanation is deeper than another along the dimension of autonomy when the explanatory connection between the explanans and explanandum is less sensitive to the breakdown of the relevant dynamical modelling frameworks or laws in regimes at very different scales from that of the explanandum. The trans-Planckian problem in cosmology can be understood as a threat to the autonomy of the explanations for key cosmic phenomena, such as the scale-invariance of the density fluctuations, based upon a breakdown in separation of scales. In this section we will consider the particular relevance of the problem to our explanatory comparison between inflationary and bouncing models. We will find that the problem gives us give reason to believe that the explanations offered by inflationary models are in general terms less deep than those offered by bouncing cosmology along the dimension of autonomy. We will also find that in the context of the so-called Trans-Planckian Censorship Conjecture (TCC), the sub-set of compatible inflationary models are such that they have greater depth in the dimension of autonomy; however, this comes at the expense of depth in the dimension of dynamical fine-tuning.         

\subsection{Inflation and Planck Scale Physics}\label{InflatingPlanck}

The trans-Planckian problem for inflationary cosmology can be stated as follows. First, we observe that scalar perturbations result from tiny fluctuations in the fields driving cosmological dynamics. For an inflating space-time, the exponential expansion present in any such scenario stretches these fluctuations exponentially. Second, we note that inflation needs to last for a minimum length of time in order to solve the horizon and flatness problems. Third, we can then reason that if inflation lasts for a sufficient duration of time, fluctuation modes that originated as trans-Planckian modes (i.e., modes that are smaller than the Planck length) can be stretched such that they exit the Hubble radius and `freeze'. These frozen modes undergo a quantum to classical transition, re-enter the horizon, and seed the scale-invariant density perturbations that form large scale structure, a cosmological explanandum of considerable importance. Thus, at least some of these \textit{classical} fluctuations originated as \textit{quantum} fluctuations smaller than the Planck length. In other words, this means that these trans-Planckian modes are described by the cosmological framework comprised of the perturbed Friedmann equations and quantum field theory, when it is clear that this lies well outside this framework's domain of validity. The problem can then be stated qualitatively in terms of a \textit{sensitive dependence} between the prediction of a scale-invariant spectrum in inflationary cosmology and \textit{hidden assumptions} about super-Planck scale physics \parencite{Martin:2001}. 

Before providing a more detailed description of the problem in terms of a concrete cosmological model, let us briefly set out the implications of the problem for the explanatory depth of inflation. First, and most obviously, the trans-Planckian problem implies that for the inflationary explanation of the scale-invariant spectrum to obtain, one needs to add supplementary conditions relating to the relevant hidden assumptions regarding the super-Planck scale physics. Most prominently, as we shall discuss shortly, this seemingly requires some assumption regarding the adiabaticity within the Planck scale initial conditions or dynamics. This explicitly sacrifices at least some degree of explanatory depth along either the initial conditions or dynamical fine-tuning dimensions. Even more problematically, such a modification to the explanation has dire consequences for the explanatory depth along the dimension of autonomy. The physical scale of the explanans makes specific reference to details of the physics at the Planck scale, which is well beyond the domain of applicability for the explanans' dynamical laws. The mismatch with the physical scale of the explanandum is then around thirty orders magnitude (using a comparison between the Planck temperature and the CMB temperature). We can thus see why the trans-Planckian problem means that the inflationary explanation for the scale-invariant spectrum is rendered shallow along the dimension of autonomy and at least somewhat shallower along the dimensions of fine-tuning (either initial conditions or dynamical depending on the formulation of the problem).    

To give a more concrete explanation of the problem we can build upon the analogy with the trans-Planckian problem in black hole thermodynamics.\footnote{This problem is subject to a detailed philosophical treatment in \parencite{gryb:2020} and in what follows we build on that discussion, in particular \S2.3 and \S4.3} Soon after Hawking's famous prediction that black holes produce thermal radiation \parencite{hawking:1975}, it was noted that the derivation of Hawking radiation makes essential use of a breakdown in the separation between micro- and macro-scales \parencite{gibbons:1977}. Following the formulation of \textcite{helfer:2003}, it can be demonstrated that modes measured as energies lower than the Planck scale (i.e. `cis-' rather than `trans-' Planckian energies) by stationary observers near future time-like infinity must have originated as trans-Planckian modes from the point of view of free-falling observers less than a Planck unit of proper time before falling through the horizon. The Hawking radiation incident on a finite, stationary detector far away from the black hole can therefore be traced back to what are, for free-falling observers, trans-Planckian energies at the horizon.\footnote{For more on the trans-Planckian problem for Hawking radiation see \parencite{unruh:1981,jacobson:1991,Jacobson:1993,unruh:1995,PhysRevD.52.4559}. Accessible introductions are \parencite[\S7]{jacobson:2005} and \parencite[pp.36-8]{harlow:2016}.} Significantly, as noted by \textcite[p.79]{jacobson:2005}, the trans-Planckian problem amounts to ``a breakdown in the usual separation of scales invoked in the application of effective field theory".\footnote{For discussion of the general connection between various senses of autonomy and EFTs see  \textcite{Crowther:2018,franklin:2020}.}
The black hole trans-Planckian problem presents a serious challenge to the prospect of a deep explanation for Hawking radiation in precisely the sense of autonomy that we have articulated. The original Hawking-style derivation connects explanans and explananda across hugely different scales in such a manner that the radiative effect to be explained is highly sensitive to assumptions regarding Planck scale physics. This renders the explanation less modally robust in the sense that it is sensitively dependent upon the detailed physics of a very different scale.

Various responses to the black hole trans-Plackian problem have been represented within the literature.\footnote{Of particular interest are arguments based upon respectively: i) the Unruh effect and equivalence principle \parencite{Agullo:2009wt}; ii) horizon symmetries \parencite{Birmingham:2001qa,PhysRevD.77.024018,Iso:2006wa}; iii) the adiabatic theorem and particular `nice slice' representation \textcite{polchinski:1995}; and iv) connections between non-thermal vacuum states and violation of the semi-classical Einstein equations \parencite{candelas1980vacuum,sciama:1981}. See  \textcite{harlow:2016,Wallace:2017a,gryb:2020} for further discussion.} Most relevant for our purposes are approaches that appeal to modified dispersion relations \parencite{unruh:2005,himemoto:2000,Barcelo:2008qe}. Here the idea is that quantum gravity corrections to the Hawking spectrum can be modelled in terms of modifications to the dispersion relation of the high-energy Hawking modes. The late-time flux of Hawking modes is explicitly computed with the modified dispersion relations using a straightforward generalisation of Hawking's original derivation. Provided the modifications to the dispersion relation satisfy a number of plausible criteria, the Hawking spectrum can be shown to be insensitive to the modifications. The thermal spectrum of radiation is thus robust against a wide variety of potential modifications to the dispersion relation and even if the modes responsible for black hole radiation do originate from the trans-Planckian regime, the thermal properties of such radiation will very likely be insensitive to such Planck scale physics. The virtue of explanations of Hawking radiation based upon modified dispersion relations is thus precisely their explanatory depth in the sense of autonomy: the explanations provided are such that the explanatory connection between the explanans and explanandum is insensitive to the breakdown of our dynamical modelling frameworks or laws in regimes at very different scales from that of the explanandum. The relevant cross-scale explanatory connection does not display a high-level sensitivity to assumptions regarding
the scale of the explanans and thus the explanation provided is deep in the sense of autonomy despite connecting very different scales.

The contrast with the cosmological trans-Planckian problem can then be explicitly made by applying a similar modified dispersion relation approach in the context of inflationary models. The key idea is to consider non-trivial relation between the physical frequency and comoving momentum of fluctuation modes \parencite{Martin:2001,Martin:2003kp,Brandenberger:2012aj}. Most straightforwardly, we can consider scalar metric fluctuations and modify the standard linear dispersion relation such that: 
\begin{equation*}
\omega^2 = k ^2 \; \rightarrow \; \omega=F(k),   
\end{equation*}
where $k \equiv \frac{n^2}{a^2}$, $n$ and $k$ are the comoving and physical wave-numbers respectively, and $F$ is assumed to be a non-linear function. 

The modifications to the dispersion relation can then be fed into the dynamics of simple inflationary models and the quantitative effects on the resulting power-spectrum studied. What can be shown is that the scale-invariance of the power spectrum depends sensitively on the form of modification. In particular, it can be shown that it is only if the modified dispersion relation satisfies an adiabaticity constraint in the UV sector that we can avoid  the spectrum of cosmological perturbations acquiring a blue tilt whose spectral slope can well exceed current limits. This amounts to a specific choice of quantum gravity dynamics that is compatible with adiabaticity. 
The modified dispersion relation approach thus directly implies that inflationary explanations for the scale-invariant spectrum are required to sacrifice explanatory depth in terms of both autonomy and the dynamical fine-tuning dimensions.\footnote{We should note here that the modified dispersion relation based arguments towards this conclusion are not entirely without controversy. See discussions of   \textcite{Kaloper:2002a,Kaloper:2002b} and \textcite{brandenberger:2002,burgess:2003}.}    

An alternative approach is to \textit{not} evolve the fluctuation modes during the time period in which their wavelength is smaller than the length scale of new physics. This corresponds to introducing a time-like `new physics hypersurface' on which special initial conditions are imposed. As noted by \textcite{Brandenberger:2012aj}, under such an approach the trans-Planckian problem has simply been shifted to the problem of choosing initial conditions on the new physics hypersurface. Furthermore, one version of this approach consists in explicitly starting modes off in their local adiabatic vacuum. Some physicists have argued that this is indeed a plausible approach to the problem and developed analogies in support of this argument with more familiar systems and their EFT descriptions \parencite{Burgess:2020nec}. Even here though, this converts the dynamical fine-tuning at the trans-Planckian scales needed in the modified dispersion relation approach to a form of initial conditions fine-tuning. Moreover, once more, such an approach will inevitably sacrifice explanatory depth along the dimension of autonomy. 

Part of the reason why this form of shallowness is particularly concerning is not only that the explanation depends on specific modelling choices at different energy scales, but also that our present knowledge of this particular energy scale is almost entirely speculative. The optimistic way of viewing the situation is that employing either of the above approaches could give us powerful hints and empirical constraints on Planck scale physics, while the pessimistic view worries that these approaches require a concerning reliance on speculation.

\subsection{Bouncing Cosmologies and Planck Scale Physics}

The relationship between various proposals for bouncing cosmology and the physics of the Planck scale is a key factor in evaluating the models. One of the primary motivations for the introduction of a bounce is `resolution' of the initial singularity.\footnote{For discussion of criteria for singularity resolution in quantum cosmology see \parencite{Thebault:2023}.} A generic feature of bouncing cosmologies is that an initially contracting phase connects us to the currently expanding one via a bounce that takes place at some minimal value of the scale factor, hence avoiding the blow-up in scalar curvature invariants generically associated with the cosmic big bang singularity \parencite{hawking:1970,ellis:1977,thorpe:1977}.\footnote{It is worth noting here the contrast with inflation where it has been shown that the Penrose-Hawking singularity theorems can be extended to show that a broad range of ‘physically reasonable’ eternal inflationary universes are necessarily inextenable and geodesically past incomplete, and therefore singular in the relevant sense \parencite{Borde:2001nh}.} 
There is thus a quite general sense in which the bouncing cosmology paradigm can be expected to provide explanations which are autonomous from the Planck scale.  

At a more specific level, in the context of the trans-Planckian problem, we can find good reasons to expect that explanatory depth along the dimension of autonomy will obtain for explananda such as the scale-invariance of the spectrum of density fluctuations. In particular, while fluctuations will shrink somewhat during a contraction phase, as long as the bounce remains far from the Planck regime, the fluctuations of interest never come close to approaching the trans-Planckian regime \parencite{Cai:2014bea, Brandenberger:2021pzy}. According to \textcite{Brandenberger:2016vhg}, if the energy scale of the bounce corresponds to the same energy scale as in typical inflation models, then the wavelengths of scales corresponding to observed cosmic microwave background anisotropies were always larger
than $1$ mm. This means that the relevant explanations can be provided in a manner such that they are autonomous from the Planck scale without requiring further dynamical or initial conditions fine-tuning.\footnote{A similar argument can run for the autonomy of the bouncing cosmological explanation of the smoothness of the universe from potential destabilisation effects of chaotic evolution in the asymptotic BKL regime. See \textcite{ijjas2018bouncing} and \textcite{Battefeld:2014uga} for detailed discussion.}  Indeed, proponents of such bouncing models argue that models in which all stages of the universe's evolution are dominated classically are advantageous in this sense because ``there is no quantum-to-classical transition to be explained" \parencite[p.9]{ijjas2018bouncing}, a significant contrast to the trans-Planckian problem in inflation. 

Another potential benefit of non-singular approaches that avoid the Planck scale is that they can offer a resolution of the \textit{entropy problem}. \textcite{Penrose:1980ge, Penrose:1988mg} has argued that any universe that emerges from a gravitational singularity would naturally be expected to be maximally entropic as all degrees of freedom should be excited (matter, radiation, gravitational, etc.). In particular, the gravitational entropy associated with tidal effects and inhomogeneities should dominate this early state and contribute an enormous amount of entropy to the universe. Furthermore, these inhomogeneities should be so significant that even inflation would seemingly be precluded from beginning at all. Yet, the universe we observe in the CMB is nearly maximal in its thermal entropy and completely negligible in its gravitational entropy, which already corresponds to a very low initial entropy state. 
This implies that there must have been an even more special initial state in the preceding inflationary epoch. A non-singular bouncing model seemingly avoids these entropy puzzles because its autonomy naturally protects it from the singularities that lead to such large expectations for the initial entropy of the universe. 

While researchers within the bouncing paradigm consider resolving this problem to be a major advantage over inflation, the explanatory comparison is a little more difficult to frame in the terms we have introduced. In part, this is because the aforementioned issues regarding probability measures mean that arguments based upon appeal to relative typicality are not well-defined  \parencite{Schiffrin:2012}. Furthermore, the inflation community has also pointed out that a full resolution of questions surrounding singularities will likely only come with a theory of quantum gravity that describes Planck scale physics \parencite{Guth:2013sya}. If one has every expectation that we can develop a theory that will address these questions, combined with the understanding that these various cosmological paradigms are effective field theories, we can see why inflation theorists are less concerned by the entropy problem and singularity avoidance. In this way, the discussion somewhat mirrors that for trans-Planckian modes, where bouncing cosmologists see an important explanatory advantage and inflationary cosmologists instead see hints about physics at higher energy scales. The difference is that quantum gravity is an appropriate and natural arena in which to investigate questions surrounding gravitational singularities, whereas there seems to be something genuinely perplexing, if not problematic, about mixing Planck scale physics with descriptions of essentially classical density perturbations associated with the large-scale structure of the universe.


\subsection{Inflation and the Trans-Planckian Censorship Conjecture} 
\label{TCC}

Let us now return our discussion to inflationary models and consider a third potential response to inflation's trans-Planckian problem: the recently formulated \textit{Trans-Planckian Censorship Conjecture (TCC)} \parencite{Brandenberger:2021pzy, Bedroya:2019snp, Bedroya:2019tba}.  The TCC holds that observers such as us are necessarily screened from trans-Planckian modes, in analogy with the Cosmic Censorship Conjecture (CCC), which, in its weak form, can be plausibly interpreted to assert that for `physically reasonable' spacetimes, there can be no singularities visible for observers at `late' times (i.e., near future null infinity) \parencite{Penrose:1969,Penrose:1973}. In both cases the idea is that there is a physical constraint that prevents observers from being exposed to radiative modes which have in their past probed arbitrarily high frequencies.   

In qualitative terms, the TCC amounts to an assertion that the trans-Planckian problem can be circumvented \textit{by fiat} such that structure formation in the early universe is autonomous with regard to the physics of the Planck regime \parencite{Schneider:2021}. In more quantitative terms, the TCC consists in the specification of a condition which enforces the autonomy of inflationary models from the Planckian scale. This condition can be expressed explicitly via the relation \parencite{Brandenberger:2021pzy}: 
\begin{equation}\label{tccUB}
    \frac{a_f}{a_i} \ell_p < \frac{1}{H_f},
\end{equation}
where inflation begins at scale factor $a_i$ and ends at scale factor $a_f$. This equation implies that a fluctuation the size of the Planck length $\ell_p$ cannot be amplified such that it is greater than the Hubble radius at end of inflation. In other words, such trans-Planckian modes are not allowed to exit the horizon and `freeze', only to re-enter the horizon as classical modes later. As long as this inequality holds, observers are protected from trans-Planckian modes.\footnote{In this context, there is a connection between the TCC and the Swampland Conjectures in string theory \parencite{Bedroya:2019snp}.}

Assuming the truth of the TCC, an inflation model will necessarily be autonomous from Planck scale physics: the explanations offered for the relevant cosmic explananda are stipulated to be such that the relevant physical scales are closely matched. We do not need to speculate about initial conditions on trans-Planckian scales in order to offer an explanation for classical large scale structure formation. Inflationary explanations with the TCC in hand are deep in the explanatory dimension of autonomy since the explanans, explananda, and domain of applicability of the relevant dynamical laws are all within the same broad arena. 

However, this success along the autonomy dimension of explanatory depth comes with an attendant cost. The inequality (\ref{tccUB}) represents an upper bound on the amount of inflation that can occur without violating the TCC; however, there is also a lower bound if inflation's dynamical, causal explanations are to function properly. The lower bound is given by the following \parencite{Brandenberger:2021pzy}:
\begin{equation}\label{tccLB}
    \frac{a_i}{a_0} \frac{1}{H_0} < \frac{1}{H_i},
\end{equation}
where $a_0$ denotes the current scale factor and $H_0^{-1}$ denotes the current Hubble radius, while $i$ denotes the beginning of inflation. The inequality (\ref{tccLB}) implies that modes that are within the horizon now must have been in causal contact (i.e., within the Hubble radius $H_i^{-1}$) at the beginning of inflation. This is necessary for inflationary dynamics to offer a causal explanation of structure formation. 

The two bounds given by (\ref{tccUB}) and (\ref{tccLB}) can be combined to constrain the energy scale of inflation, such that inflation would have had to occur at $\sim 10^8$GeV, or several orders of magnitude lower than the GUT scale ($\sim 10^{15}$GeV) that inflation has traditionally been believed to operate within. This has significant implications for inflationary dynamics. Among other things, it implies that the inflaton potential must be dynamically fine-tuned in order to match the observed amplitude of scalar fluctuations, while also operating within these constraints on the energy scale \parencite{Brandenberger:2021pzy, Bedroya:2019tba}. It should also be noted, as \textcite{Brandenberger:2021pzy} acknowledges, that TCC constraints on inflation are weaker in more complicated multi-field inflation models. As these types of models are beyond the scope of this paper, we will not address them in detail here. However, this is relevant because some cosmologists who favour inflation believe that more realistic models of inflation could plausibly feature multiple fields and multiple stages of inflation. As \textcite{Guth:2013sya} emphasize, recent developments in high energy physics and our current understanding of string vacua point towards the idea that realistic models of inflation may indeed be significantly more intricate than the standard single-field picture. Our analysis also applies in this case, just with the caveat that constraints coming from the TCC, along with the fine-tuning needed to operate within them, would need to be revised accordingly.

The implication, within our account for evaluating depth of explanations, is that inflation with the TCC trades explanatory shallowness in terms of autonomy for explanatory shallowness in terms of dynamical fine-tuning. The problem of choosing between inflationary explanations with and without the TCC, like that of choosing between inflation and bouncing explanations, then becomes one of \textit{weighting} dimensions of depth. We will consider this issue and its broader implications for both cosmology and the nature of scientific methodology in the final section. 

\section{Dimensions of Depth and Heuristics}

\noindent In this paper we have understood explanatory depth as a non-unitary concept with different dimensions relevant to different domains. The domain of primordial cosmology is one in which the three most relevant dimensions can be understood as i) initial conditions fine-tuning; ii) dynamical fine-tuning; and iii) autonomy. Following the insightful analysis of \textcite{Azhar:2021}, we diagnosed the explanatory preference of contemporary cosmologists for the inflationary paradigm over the HBB paradigm as being based upon the greater explanatory depth along the dimension of initial conditions fine-tuning. This observation  encodes a primarily descriptive rational reconstruction of the preference of cosmologist for inflation over the HBB. 

Where things become more complex, and our account starts to blend the normative and descriptive, is in the explanatory comparison between inflationary and bouncing paradigms. In that context, we have isolated what we take to be the principal factor motivating the explanatory preference of most, although not all, cosmologists for the inflationary approach. Both paradigms successfully provide explanations that avoid initial conditions fine-tuning and offer far more depth than the HBB model which preceded them. However, the paradigms can be differentiated along the dimension of dynamical fine-tuning. Due to the need to avoid unphysical instabilities, models within the bouncing paradigm can be understood to display a form of dynamical fine-tuning which renders the relevant explanations lacking in depth along this dimension. That is, the explanatory relationship between explanans and explanandum is highly sensitive to variations in the dynamical structures \textit{because} the physical viability of such models requires a significant degree of dynamical fine-tuning.

Taken on its own, dynamical fine-tuning allows us to appreciate why most theorists favour an inflationary account of the early universe. However, things become more controversial when we consider the explanatory dimension of autonomy. In this context, the trans-Planckian problem afflicts inflationary models, but not bouncing models, and represents a severe challenge to the autonomy of the relevant explanations, particularly with regard to the scale-invariance of the power spectrum. This bifurcates the explanatory merits of inflation into two different routes. 

One possibility is that we accept that inflation's explanatory merits are shallow along this dimension of autonomy. In this case, inflation could potentially provide invaluable access to trans-Planckian physics and quantum gravity. This reflects an exciting opportunity where ``unobservable physics might unexpectedly come within observational reach" \parencite[p.1]{Burgess:2020nec}. However, in this scenario, inflation ends up being a bridesmaid rather than \textit{the} bride, in that it stands adjacent to the trans-Planckian physics that is also responsible for the salient features of the observable universe. This is of course a trade-off that many physicists are happy to make, but in this case we must acknowledge that inflation does not carry the same explanatory weight most often attributed to it as much of this explanatory burden is then shifted to the relevant trans-Planckian details. 

The other possibility is to make a move to restore the autonomy of inflation's explanatory power. The trans-Planckian problem can be ameliorated by appeal to the trans-Planckian Censorship Conjecture. Such a move, in turn, then requires inflationary models to be themselves dynamically fine-tuned in a non-trivial way so as to avoid violating the conjecture. The result of this move is that both inflationary and bouncing models display a lack of explanatory depth in terms of dynamical fine-tuning. The main source of the difference between the paradigms is then that bouncing models need to be dynamically fine-tuned to be viable in principle, whereas inflation models need to be dynamically fine-tuned in order for the desired explanatory relationships to hold. Descriptively, it does seem like the dynamical fine-tuning evident in bouncing models is judged more harshly because it concerns the physical viability of the model, rather than the dynamical fine-tuning invoked to match observational constraints. The full situation can be concisely represented in Table \ref{tab:DDtable}.  




\begin{table}[]

\begin{tabular}{lcccc}
\multicolumn{1}{c}{Dimension of Depth}  & \begin{tabular}[c]{@{}c@{}}Inflation\\ (no TCC)\end{tabular} & \begin{tabular}[c]{@{}c@{}}Inflation\\ (with TCC)\end{tabular} & Big Bounce   \\ 
\cline{2-4} 
\multicolumn{1}{l|}{Initial conditions fine-tuning} & \multicolumn{1}{c|}{\cmark/\xmark}  & \multicolumn{1}{c|}{\cmark}  & \multicolumn{1}{c|}{\cmark}  \\ \cline{2-4} 
\multicolumn{1}{l|}{Dynamical fine-tuning}   & \multicolumn{1}{c|}{\xmark/\cmark}  & \multicolumn{1}{c|}{\xmark} & \multicolumn{1}{c|}{\xmark} \\ 
\cline{2-4} 
\multicolumn{1}{l|}{Autonomy}  & \multicolumn{1}{c|}{\xmark}  & \multicolumn{1}{c|}{\cmark} & \multicolumn{1}{c|}{\cmark}  \\ 
\cline{2-4} 
\end{tabular}
\bigskip

\caption{Deeper explanations are marked by a \cmark\:and correspond to \textit{less} fine-tuning. The \xmark/\cmark\:in Inflation (no TCC) reflects a choice with regard to how to avoid the trans-Planckian problem. }
\label{tab:DDtable}
\end{table}


Where does this leave us? One the one hand, our analysis provides a degree of clarity with regard to the reasons why cosmologists so strongly disagree with regard to the extra-empirical merits of the various paradigms: on our account they may simply be arguing at cross purposes by relying upon comparisons along incommensurable dimensions of depth. On the other hand, the result of this explanatory incommensurability is to blunt the normative utility of our analysis so far as we would like to provide a means through which to recommend scientists towards the deepest explanation available. On our analysis, there is no fact of the matter with regard to whether the explanations provided by inflationary or bouncing paradigms are deeper because there are multiple relevant dimensions of depth without a common measure of comparison. 

What we would like to propose, on a more constructive note, is that the explanatory preference with regard to the different dimensions of depth can be understood in terms of differing attitudes with regard to heuristics for future model building. In particular, the reason why explanations that lack depth qua initial conditions fine-tuning are so unsatisfactory is, at least in part, due to the heuristic sterility of explanations of phenomena that appeal to special initial conditions.\footnote{Here we would similarly categorise explanations for temporal asymmetry that rely on the so-called past hypothesis \textcite{Earman:2006,Gryb:2021}.} 

The choice between explanations that are deeper along the dimensions of autonomy and dynamical fine-tuning might be similarly framed in terms of their respective forms of heuristic fecundity. The heuristic value of an autonomous but dynamically fine-tuned explanation can be understood in terms of the positive heuristics provided for theoretical model building in a constrained space within limitations on both the realm of relevant empirical phenomena and the possible dynamical structures that can be implements. By contrast, the value of an explanatory approach that is deep in virtue of not being dynamically fine-tuned, but shallower in virtue of lack of autonomy, might be understood in broadly empiricist terms: the failure of autonomy opens a window for plausible empirical constraints connecting vastly different energy scales, c.f.\ \textcite{Schneider:2021}. In this sense, trade-offs between dimensions of explanatory depth might be interpreted as encoding differing methodological stances rather than a choice between strictly incommensurable alternatives.

\section*{Acknowledgments}

\noindent We are deeply indebted to Robert Brandenberger, Matt Davies, Casey McCoy, Richard Dawid, Alex Franklin, James Read, Chris Smeenk, Mike Schneider, Paul Steinhardt, and two anonymous referees for extremely helpful comments of a draft version of this paper. We are also very grateful for suggestions and comments from an audience in Stockholm. Work on this project was supported through a British Academy Mid-Career grant. 

\printbibliography

@article{Thebault:2023,
  title={Big bang singularity resolution in quantum cosmology},
  author={Th{\'e}bault, Karim P Y},
  journal={Classical and Quantum Gravity},
  volume={40},
  number={5},
  pages={055007},
  year={2023},
  publisher={IOP Publishing}
}

@article{Cretu:2022,
  title={Authentication, scale-relativity, and relational kindhood},
  author={Cre{\c{t}}u, Ana-Maria},
  journal={Synthese},
  volume={200},
  number={1},
  pages={1--21},
  year={2022},
  publisher={Springer}
}

@book{Ladyman:2020,
  title={What is a complex system?},
  author={Ladyman, James and Wiesner, Karoline},
  year={2020},
  publisher={Yale University Press}
}

@book{Crowther:2018,
  title={Effective spacetime},
  author={Crowther, Karen},
  year={2018},
  publisher={Springer}
}

@article{castellani:2022,
  title={Renormalization group methods: Which kind of explanation?},
  author={Castellani, Elena and Margoni, Emilia},
  journal={Studies in History and Philosophy of Science},
  volume={95},
  pages={158--166},
  year={2022},
  publisher={Elsevier}
}

@article{reutlinger:2017,
  title={Are causal facts really explanatorily emergent? Ladyman and Ross on higher-level causal facts and renormalization group explanation},
  author={Reutlinger, Alexander},
  journal={Synthese},
  volume={194},
  number={7},
  pages={2291--2305},
  year={2017},
  publisher={Springer}
}

@incollection{Darrigol:2013,
    author = {Darrigol, Olivier},
    isbn = {9780195392043},
    title = "{For a Philosophy of
            Hydrodynamics}",
    booktitle = "{The Oxford Handbook of Philosophy of Physics}",
    publisher = {Oxford University Press},
    year = {2013},
    %month = {02},
    %abstract = "{This chapter discusses the need for a philosophy of hydrodynamics and the lessons that can be learned from the historical development of fluid mechanics. It explains that hydrodynamics has been not given attention by philosophers of physics because of a lack of detailed historical studies of hydrodynamics, and highlights the need for idealizations and modeling strategies for this theory to be applicable. The chapter also considers the structures of phenomenological theories and the so-called modular structure of hydrodynamics.}",
    doi = {10.1093/oxfordhb/9780195392043.013.0002},
    %url = {https://doi.org/10.1093/oxfordhb/9780195392043.013.0002},
    %eprint = {https://academic.oup.com/book/0/chapter/215026050/chapter-ag-pdf/44587526/book\_28318\_section\_215026050.ag.pdf},
}

@inproceedings{Linde:2014nna,
    author = "Linde, Andrei",
    title = "{Inflationary Cosmology after Planck 2013}",
    booktitle = "{100e Ecole d'Ete de Physique: Post-Planck Cosmology}",
    eprint = "1402.0526",
    archivePrefix = "arXiv",
    primaryClass = "hep-th",
    doi = "10.1093/acprof:oso/9780198728856.003.0006",
    pages = "231--316",
    year = "2015"
}

@article{franklin:2020,
  title={Whence the effectiveness of effective field theories?},
  author={Franklin, Alexander},
  journal={The British Journal for the Philosophy of Science},
  year={2020},
  publisher={The University of Chicago Press}
}

@article{Darrigol:2009,
  title={A simplified genesis of quantum mechanics},
  author={Darrigol, Olivier},
  journal={Studies in History and Philosophy of Science Part B: Studies in History and Philosophy of Modern Physics},
  volume={40},
  number={2},
  pages={151--166},
  year={2009},
  publisher={Elsevier}
}

@book{Palacios:2022,
  title={Emergence and Reduction in Physics},
  author={Palacios, Patricia},
  year={2022},
  publisher={Cambridge University Press}
}

@book{Broer:2011,
  title={Dynamical systems and chaos},
  author={Broer, Hendrik Wolter and Takens, Floris},
  volume={172},
  year={2011},
  publisher={Springer}
}

@article{Wallace:2017a,
	Author = {Wallace, David},
	Journal = {Studies in History and Philosophy of Science Part B: Studies in History and Philosophy of Modern Physics},
	Pages = {52--67},
	Publisher = {Elsevier},
	Title = {The case for black hole thermodynamics part I: Phenomenological thermodynamics},
	Volume = {64},
	Year = {2018}}

@article{Martin:2001,
	Author = {J{\'e}r{\^o}me Martin and Robert H. Brandenberger},
	Doi = {10.1103/PhysRevD.63.123501},
	Journal = {Physical Review D},
	Number = {12},
	Title = {Trans-Planckian problem of inflationary cosmology},
	Volume = {63},
	Year = {2001}}

@article{sciama:1981,
	Author = {Sciama, Dennis W and Candelas, P and Deutsch, D},
	Journal = {Advances in Physics},
	Number = {3},
	Pages = {327--366},
	Publisher = {Taylor \& Francis},
	Title = {Quantum field theory, horizons and thermodynamics},
	Volume = {30},
	Year = {1981}}

@article{saatsi:2018,
	Author = {Saatsi, Juha and Reutlinger, Alexander},
	Journal = {Philosophy of Science},
	Number = {3},
	Pages = {455--482},
	Publisher = {University of Chicago Press Chicago, IL},
	Title = {Taking Reductionism to the Limit: How to Rebut the Antireductionist Argument from Infinite Limits},
	Volume = {85},
	Year = {2018}}

@article{candelas1980vacuum,
	Author = {Candelas, Philip},
	Journal = {Physical Review D},
	Number = {8},
	Pages = {2185},
	Publisher = {APS},
	Title = {Vacuum polarization in Schwarzschild spacetime},
	Volume = {21},
	Year = {1980}}

@article{helfer:2003,
	Author = {Helfer, Adam D},
	Journal = {Reports on Progress in Physics},
	Number = {6},
	Pages = {943},
	Publisher = {IOP Publishing},
	Title = {Do black holes radiate?},
	Volume = {66},
	Year = {2003}}

@incollection{gibbons:1977,
	Author = {Gibbons, GW},
	Booktitle = {Proceedings of the First Marcel Grossmann Meeting on General Relativity},
	Editor = {R Ruffini},
	Pages = {pp 449--58},
	Publisher = {North-Holland, Amsterdam},
	Title = {Quantum Processing Near Black Holes},
	Year = {1977}}

@article{harlow:2016,
	Author = {Harlow, Daniel},
	Journal = {Reviews of Modern Physics},
	Number = {1},
	Pages = {015002},
	Publisher = {APS},
	Title = {Jerusalem lectures on black holes and quantum information},
	Volume = {88},
	Year = {2016}}

@inproceedings{polchinski:1995,
    author = "Polchinski, Joseph",
    title = "{String theory and black hole complementarity}",
    booktitle = "{STRINGS 95: Future Perspectives in String Theory}",
    eprint = "hep-th/9507094",
    archivePrefix = "arXiv",
    reportNumber = "NSF-ITP-95-63",
    pages = "417--426",
    month = "7",
    year = "1995"
}

@article{PhysRevD.52.4559,
	Author = {Brout, R. and Massar, S. and Parentani, R. and Spindel, Ph.},
	Doi = {10.1103/PhysRevD.52.4559},
	Issue = {8},
	Journal = {Phys. Rev. D},
	%Month = {Oct},
	Numpages = {0},
	Pages = {4559--4568},
	Publisher = {American Physical Society},
	Title = {Hawking radiation without trans-Planckian frequencies},
	Url = {https://link.aps.org/doi/10.1103/PhysRevD.52.4559},
	Volume = {52},
	Year = {1995},
	Bdsk-Url-1 = {https://link.aps.org/doi/10.1103/PhysRevD.52.4559},
	Bdsk-Url-2 = {https://dx.doi.org/10.1103/PhysRevD.52.4559}}

@article{PhysRevD.77.024018,
	Author = {Banerjee, Rabin and Kulkarni, Shailesh},
	Doi = {10.1103/PhysRevD.77.024018},
	Issue = {2},
	Journal = {Phys. Rev. D},
	%Month = {Jan},
	Numpages = {5},
	Pages = {024018},
	Publisher = {American Physical Society},
	Title = {Hawking radiation and covariant anomalies},
	Url = {https://link.aps.org/doi/10.1103/PhysRevD.77.024018},
	Volume = {77},
	Year = {2008},
	Bdsk-Url-1 = {https://link.aps.org/doi/10.1103/PhysRevD.77.024018},
	Bdsk-Url-2 = {https://dx.doi.org/10.1103/PhysRevD.77.024018}}

@article{Iso:2006wa,
	Archiveprefix = {arXiv},
	Author = {Iso, Satoshi and Umetsu, Hiroshi and Wilczek, Frank},
	Doi = {10.1103/PhysRevLett.96.151302},
	Eprint = {hep-th/0602146},
	Journal = {Phys. Rev. Lett.},
	Pages = {151302},
	Primaryclass = {hep-th},
	Reportnumber = {MIT-CTP-3714, KEK-TH-1062, OIQP-05-23},
	Slaccitation = {%%CITATION = HEP-TH/0602146;%%},
	Title = {{Hawking radiation from charged black holes via gauge and gravitational anomalies}},
	Volume = {96},
	Year = {2006}}

@article{Birmingham:2001qa,
	Archiveprefix = {arXiv},
	Author = {Birmingham, Danny and Gupta, Kumar S. and Sen, Siddhartha},
	Doi = {10.1016/S0370-2693(01)00354-9},
	Eprint = {hep-th/0102051},
	Journal = {Phys. Lett.},
	Pages = {191-196},
	Primaryclass = {hep-th},
	Reportnumber = {SINP-TNP-01-02},
	Slaccitation = {%%CITATION = HEP-TH/0102051;%%},
	Title = {{Near horizon conformal structure of black holes}},
	Volume = {B505},
	Year = {2001}}

@article{Barcelo:2008qe,
	Archiveprefix = {arXiv},
	Author = {Barcelo, C. and Garay, L. J. and Jannes, G.},
	Doi = {10.1103/PhysRevD.79.024016},
	Eprint = {0807.4147},
	Journal = {Phys. Rev.},
	Pages = {024016},
	Primaryclass = {gr-qc},
	Slaccitation = {%%CITATION = ARXIV:0807.4147;%%},
	Title = {{Sensitivity of Hawking radiation to superluminal dispersion relations}},
	Volume = {D79},
	Year = {2009}}

@article{Agullo:2009wt,
	Archiveprefix = {arXiv},
	Author = {Agullo, Ivan and Navarro-Salas, Jose and Olmo, Gonzalo J. and Parker, Leonard},
	Doi = {10.1103/PhysRevD.80.047503},
	Eprint = {0906.5315},
	Journal = {Phys. Rev.},
	Pages = {047503},
	Primaryclass = {gr-qc},
	Slaccitation = {%%CITATION = ARXIV:0906.5315;%%},
	Title = {{Insensitivity of Hawking radiation to an invariant Planck-scale cutoff}},
	Volume = {D80},
	Year = {2009}}

@article{Reutlinger2014-REUWIT,
	journal = {Philosophy of Science},
	doi = {10.1086/677887},
	pages = {1157--1170},
	volume = {81},
	year = {2014},
	number = {5},
	title = {Why Is There Universal Macrobehavior? Renormalization Group Explanation as Noncausal Explanation},
	author = {Alexander Reutlinger}
}

@book{batterman:2002,
	Author = {Batterman, Robert W},
	Publisher = {Oxford: Oxford University Press},
	Title = {The devil in the details: Asymptotic reasoning in explanation, reduction, and emergence},
	Year = {2002}}

@article{batterman:2000,
	Author = {Batterman, Robert W},
	Journal = {The British Journal for the Philosophy of Science},
	Number = {1},
	Pages = {pp. 115--45},
	Publisher = {Br Soc Philosophy Sci},
	Title = {{Multiple realizability and universality}},
	Volume = {\textbf{51}},
	Year = {2000}}

@article{himemoto:2000,
	Author = {Himemoto, Yoshiaki and Tanaka, Takahiro},
	Journal = {Physical Review D},
	Number = {6},
	Pages = {p. 064004},
	Publisher = {APS},
	Title = {{Generalization of the model of Hawking radiation with modified high frequency dispersion relation}},
	Volume = {\textbf{61}},
	Year = {2000}}

@article{Jacobson:1993,
	Author = {Ted Jacobson},
	Doi = {10.1103/PhysRevD.48.728},
	Journal = {Physical Review D},
	Number = {2},
	Pages = {728--41},
	Title = {{Black hole radiation in the presence of a short distance cutoff}},
	Volume = {48},
	Year = {1993}}

@article{jacobson:1991,
	Author = {Jacobson, Theodore},
	Journal = {Physical Review D},
	Number = {6},
	Pages = {1731},
	Publisher = {APS},
	Title = {{Black-hole evaporation and ultrashort distances}},
	Volume = {44},
	Year = {1991}}

@article{unruh:1995,
	Author = {Unruh, WG},
	Journal = {Physical Review D},
	Number = {6},
	Pages = {p. 2827},
	Publisher = {APS},
	Title = {{Sonic analogue of black holes and the effects of high frequencies on black hole evaporation}},
	Volume = {\textbf{51}},
	Year = {1995}}

@article{unruh:2005,
	Author = {Unruh, William G and Sch{\"u}tzhold, Ralf},
	Journal = {Physical Review D},
	Number = {2},
	Pages = {024028},
	Publisher = {APS},
	Title = {{Universality of the Hawking effect}},
	Volume = {71},
	Year = {2005}}

@incollection{jacobson:2005,
	Author = {Jacobson, Ted},
	Booktitle = {Lectures on Quantum Gravity},
	Pages = {pp. 39--89},
	Publisher = {New York: Springer},
	Title = {{Introduction to quantum fields in curved spacetime and the Hawking effect}},
	Year = {2005}}

@article{hawking:1975,
	Author = {Hawking, Stephen W},
	Journal = {Communications in mathematical physics},
	Number = {3},
	Pages = {199--220},
	Publisher = {Springer},
	Title = {{Particle creation by black holes}},
	Volume = {43},
	Year = {1975}}

@article{unruh:1981,
	Author = {Unruh, WG},
	Journal = {Physical Review Letters},
	Number = {21},
	Pages = {1351--53},
	Publisher = {APS},
	Title = {{Experimental black-hole evaporation?}},
	Volume = {46},
	Year = {1981}}

@article{franklin:2018,
  title={On the renormalization group explanation of universality},
  author={Franklin, Alexander},
  journal={Philosophy of Science},
  volume={85},
  number={2},
  pages={225--248},
  year={2018},
  publisher={Cambridge University Press}
}

@article{Smeenk:2014,
  title={Predictability crisis in early universe cosmology},
  author={Smeenk, Chris},
  journal={Studies in History and Philosophy of Science Part B: Studies in History and Philosophy of Modern Physics},
  volume={46},
  pages={122--133},
  year={2014},
  publisher={Elsevier}
}

@article{Holman:2018,
  title={How problematic is the near-Euclidean spatial geometry of the large-scale universe?},
  author={Holman, Marc},
  journal={Foundations of Physics},
  volume={48},
  number={11},
  pages={1617--1647},
  year={2018},
  publisher={Springer}
}

@article{burgess:2003,
  title={Are inflationary predictions sensitive to very high energy physics?},
  author={Burgess, Clifford P and Cline, James M and Lemieux, Fran{\c{c}}ois and Holman, Richard},
  journal={Journal of High Energy Physics},
  volume={2003},
  number={02},
  pages={048},
  year={2003},
  publisher={IOP Publishing}
}

@article{brandenberger:2002,
  title={On signatures of short distance physics in the cosmic microwave background},
  author={Brandenberger, Robert H and Martin, Jerome},
  journal={International Journal of Modern Physics A},
  volume={17},
  number={25},
  pages={3663--3680},
  year={2002},
  publisher={World Scientific}
}

@article{Kaloper:2002a,
  title={Signatures of short distance physics in the cosmic microwave background},
  author={Kaloper, Nemanja and Kleban, Matthew and Lawrence, Albion and Shenker, Stephen},
  journal={Physical Review D},
  volume={66},
  number={12},
  pages={123510},
  year={2002},
  publisher={APS}
}

@article{Kaloper:2002b,
  title={Initial conditions for inflation},
  author={Kaloper, Nemanja and Kleban, Matthew and Lawrence, Albion and Shenker, Stephen and Susskind, Leonard},
  journal={Journal of High Energy Physics},
  volume={2002},
  number={11},
  pages={037},
  year={2003},
  publisher={IOP Publishing}
}

@incollection{Smeenk:2005,
  title={False vacuum: Early universe cosmology and the development of inflation},
  author={Smeenk, Chris},
  booktitle={The universe of general relativity},
  pages={223--257},
  year={2005},
  publisher={Springer}
}

@article{Earman:2006,
  title={The “past hypothesis”: Not even false},
  author={Earman, John},
  journal={Studies in History and Philosophy of Science Part B: Studies in History and Philosophy of Modern Physics},
  volume={37},
  number={3},
  pages={399--430},
  year={2006},
  publisher={Elsevier}
}

@article{Penrose:1969,
  title={Gravitational collapse: The role of general relativity},
  author={Penrose, Roger},
  journal={Nuovo Cimento Rivista Serie},
  volume={1},
  pages={252},
  year={1969}
}

@article{Penrose:1988mg,
    author = "Penrose, R.",
    editor = "Fenyves, E. J.",
    title = "{Difficulties with inflationary cosmology}",
    doi = "10.1111/j.1749-6632.1989.tb50513.x",
    journal = "Annals N. Y. Acad. Sci.",
    volume = "571",
    pages = "249--264",
    year = "1989"
}

@INPROCEEDINGS{Penrose:1980ge,
       author = {{Penrose}, R.},
        title = "{Singularities and time-asymmetry.}",
    booktitle = {General Relativity: An Einstein centenary survey},
         year = 1979,
       editor = {{Hawking}, S.~W. and {Israel}, W.},
        month = jan,
        pages = {581-638},
       adsurl = {https://ui.adsabs.harvard.edu/abs/1979grec.conf..581P}
}

@article{Penrose:1973,
  title={Naked singularities},
  author={Penrose, Roger},
  journal={Annals of the New York Academy of Sciences},
  volume={224},
  number={1},
  pages={125--134},
  year={1973},
  publisher={Wiley Online Library}
}

@article{Guth:2013sya,
    author = "Guth, Alan H. and Kaiser, David I. and Nomura, Yasunori",
    title = "{Inflationary paradigm after Planck 2013}",
    eprint = "1312.7619",
    archivePrefix = "arXiv",
    primaryClass = "astro-ph.CO",
    reportNumber = "MIT-CTP-4490, UCB-PTH-13-08",
    doi = "10.1016/j.physletb.2014.03.020",
    journal = "Phys. Lett. B",
    volume = "733",
    pages = "112--119",
    year = "2014"
}

@article{Mukhanov:1981xt,
    author = "Mukhanov, Viatcheslav F. and Chibisov, G. V.",
    title = "{Quantum Fluctuations and a Nonsingular Universe}",
    journal = "JETP Lett.",
    volume = "33",
    pages = "532--535",
    year = "1981"
}

@article{Press_1980,
	doi = {10.1088/0031-8949/21/5/021},
	url = {https://doi.org/10.1088/0031-8949/21/5/021},
	year = 1980,
	%month = {jan},
	publisher = {{IOP} Publishing},
	volume = {21},
	number = {5},
	pages = {702--707},
	author = {William H Press},
	title = {Spontaneous Production of the Zel{\textquotesingle}dovich Spectrum of Cosmological Fluctuations},
	journal = {Physica Scripta},
	abstract = {The spectrum of initial cosmological perturbations posited by Zel'dovich and co-workers for the "pancake theory" of galaxy formation has (i) adiabatic perturbations only, and (ii) constant perturbation amplitude on all scales at their respective horizon times. Assumption (i) has recently been shown to follow from grand unified gauge theories. This paper shows that assumption (ii) may also be a consequence of spontaneous symmetry-breaking at sufficiently high temperatures. In this case the Universe may have been exactly Friedmannian originally. In a classical, U(1) Abelian Higgs model, the scale-independent dimensionless size of perturbations is here calculated to be δT00/T00 ∼ (8π/3)(Tc/m*)2 where m* is the Planck mass (1019 GeV) and Tc is the critical temperature of spontaneous symmetry-breaking, related to the mass of the gauge boson of the symmetry group. For the pancake model, one needs Tc ∼ 1017 GeV.}
}

@article{Brandenberger:2011eq,
    author = "Brandenberger, Robert H.",
    title = "{Is the spectrum of gravitational waves the \textquotedblleft{}Holy Grail\textquotedblright{} of inflation?}",
    eprint = "1104.3581",
    archivePrefix = "arXiv",
    primaryClass = "astro-ph.CO",
    doi = "10.1140/epjc/s10052-019-6883-4",
    journal = "Eur. Phys. J. C",
    volume = "79",
    number = "5",
    pages = "387",
    year = "2019"
}

@article{Brandenberger:2008nx,
    author = "Brandenberger, Robert H.",
    title = "{String Gas Cosmology}",
    eprint = "0808.0746",
    archivePrefix = "arXiv",
    primaryClass = "hep-th",
    month = "8",
    year = "2008"
}

@article{Ellis:2002we,
    author = "Ellis, George F. R. and Maartens, Roy",
    title = "{The emergent universe: Inflationary cosmology with no singularity}",
    eprint = "gr-qc/0211082",
    archivePrefix = "arXiv",
    doi = "10.1088/0264-9381/21/1/015",
    journal = "Class. Quant. Grav.",
    volume = "21",
    pages = "223--232",
    year = "2004"
}

@article{hawking:1970,
	Author = {Hawking, Stephen William and Penrose, Roger},
	Journal = {Proceedings of the Royal Society of London. A. Mathematical and Physical Sciences},
	Number = {1519},
	Pages = {529--548},
	Publisher = {The Royal Society London},
	Title = {The singularities of gravitational collapse and cosmology},
	Volume = {314},
	Year = {1970}}

@article{ellis:1977,
	Author = {Ellis, George FR and Schmidt, Bernd G},
	Journal = {General Relativity and Gravitation},
	Number = {11},
	Pages = {915--953},
	Publisher = {Springer},
	Title = {Singular space-times},
	Volume = {8},
	Year = {1977}}

@article{thorpe:1977,
	Author = {Thorpe, John A},
	Journal = {Journal of Mathematical Physics},
	Number = {5},
	Pages = {960--964},
	Publisher = {American Institute of Physics},
	Title = {Curvature invariants and space--time singularities},
	Volume = {18},
	Year = {1977}}

@article{Smeenk:2017,
  title={Philosophy of Cosmology},
  author={Smeenk, C and Ellis, G},
  journal={Stanford Encyclopedia of Philosophy},
  year={2017},
  publisher={}
}

@article{Mccoy:2020,
	author = {C. D. McCoy},
	volume = {70},
	year = {2019},
	number = {4},
	doi = {10.1093/bjps/axy014},
	journal = {British Journal for the Philosophy of Science},
	pages = {1003--1028},
	title = {Epistemic Justification and Methodological Luck in Inflationary Cosmology}
}

@article{Mccoy2017,
  title={Can typicality arguments dissolve cosmology’s flatness problem?},
  author={McCoy, CD},
  journal={Philosophy of Science},
  volume={84},
  number={5},
  pages={1239--1252},
  year={2017},
  publisher={Cambridge University Press}
}

@article{Guth:2017,
  title={A cosmic controversy},
  author={Guth, Alan H and Kaiser, David I and Linde, Andrei D and Nomura, Yasunori and Bennett, Charles L and Bond, J Richard and Bouchet, Fran{\c{c}}ois and Carroll, Sean and Efstathiou, George and Hawking, Stephen and others},
  journal={Scientific American},
  volume={317},
  number={1},
  pages={5--7},
  year={2017},
  publisher={JSTOR}
}

@article{Burgess:2020nec,
    author = "Burgess, C. P. and de Alwis, S. P. and Quevedo, F.",
    title = "{Cosmological Trans-Planckian Conjectures are not Effective}",
    eprint = "2011.03069",
    archivePrefix = "arXiv",
    primaryClass = "hep-th",
    doi = "10.1088/1475-7516/2021/05/037",
    journal = "JCAP",
    volume = "05",
    pages = "037",
    year = "2021"
}

@article{Brandenberger:2016uzh,
    author = "Brandenberger, Robert",
    title = "{Initial conditions for inflation \textemdash{} A short review}",
    eprint = "1601.01918",
    archivePrefix = "arXiv",
    primaryClass = "hep-th",
    doi = "10.1142/S0218271817400028",
    journal = "Int. J. Mod. Phys. D",
    volume = "26",
    number = "01",
    pages = "1740002",
    year = "2016"
}

@book{Mukhanov:2005sc,
    author = "Mukhanov, V.",
    title = "{Physical Foundations of Cosmology}",
    doi = "10.1017/CBO9780511790553",
    isbn = "978-0-521-56398-7",
    publisher = "Cambridge University Press",
    address = "Oxford",
    year = "2005"
}

@article{East:2015ggf,
    author = "East, William E. and Kleban, Matthew and Linde, Andrei and Senatore, Leonardo",
    title = "{Beginning inflation in an inhomogeneous universe}",
    eprint = "1511.05143",
    archivePrefix = "arXiv",
    primaryClass = "hep-th",
    doi = "10.1088/1475-7516/2016/09/010",
    journal = "JCAP",
    volume = "09",
    pages = "010",
    year = "2016"
}

@article{Ellis:2014,
  title={On the philosophy of cosmology},
  author={Ellis, George Francis Rayner},
  journal={Studies in History and Philosophy of Science Part B: Studies in History and Philosophy of Modern Physics},
  volume={46},
  pages={5--23},
  year={2014},
  publisher={Elsevier}
}

@book{Chamcham:2017,
  title={The Philosophy of Cosmology},
  author={Chamcham, Khalil and Silk, Joseph and Barrow, John D and Saunders, Simon},
  year={2017},
  publisher={Cambridge University Press}
}

@article{Gryb:2021,
  title={New difficulties for the past hypothesis},
  author={Gryb, Sean},
  journal={Philosophy of Science},
  volume={88},
  number={3},
  pages={511--532},
  year={2021},
  publisher={Cambridge University Press}
}

@article{Curiel:2015,
    author = "Curiel, Erik",
    title = "{Measure, Topology and Probabilistic Reasoning in Cosmology}",
    eprint = "1509.01878",
    archivePrefix = "arXiv",
    primaryClass = "gr-qc",
    month = "9",
    year = "2015"
}

@article{Schneider:2021,
  title={Trans-Planckian philosophy of cosmology},
  author={Schneider, Mike D},
  journal={Studies in History and Philosophy of Science Part A},
  volume={90},
  pages={184--193},
  year={2021},
  publisher={Elsevier}
}

@article{Earman:1999,
	Author = {Earman, John and Mosterin, Jesus},
	Journal = {Philosophy of Science},
	Number = {1},
	Pages = {1--49},
	Publisher = {Cambridge University Press},
	Title = {A critical look at inflationary cosmology},
	Volume = {66},
	Year = {1999}}

@article{Hitchcock:2003,
	Author = {Hitchcock, Christopher and Woodward, James},
	Journal = {No{\^u}s},
	Number = {2},
	Pages = {181--199},
	Publisher = {JSTOR},
	Title = {Explanatory generalizations, part II: Plumbing explanatory depth},
	Volume = {37},
	Year = {2003}}

@article{Jackson:1992,
	Author = {Jackson, Frank and Pettit, Philip},
	Journal = {Economics \& Philosophy},
	Number = {1},
	Pages = {1--21},
	Publisher = {Cambridge University Press},
	Title = {In defense of explanatory ecumenism},
	Volume = {8},
	Year = {1992}}

@article{Ylikoski:2010,
	Author = {Ylikoski, Petri and Kuorikoski, Jaakko},
	Journal = {Philosophical studies},
	Number = {2},
	Pages = {201--219},
	Publisher = {Springer},
	Title = {Dissecting explanatory power},
	Volume = {148},
	Year = {2010}}

@article{Weslake:2010,
	Author = {Weslake, Brad},
	Journal = {Philosophy of Science},
	Number = {2},
	Pages = {273--294},
	Publisher = {The University of Chicago Press},
	Title = {Explanatory depth},
	Volume = {77},
	Year = {2010}}

@article{gryb:2020,
    author = "Gryb, Sean and Palacios, Patricia and Th\'ebault, Karim P. Y.",
    title = "{On the Universality of Hawking Radiation}",
    eprint = "1812.07078",
    archivePrefix = "arXiv",
    primaryClass = "physics.hist-ph",
    doi = "10.1093/bjps/axz025",
    journal = "Brit. J. Phil. Sci.",
    volume = "72",
    number = "3",
    pages = "809--837",
    year = "2021"
}

@article{Batterman:2018,
	Author = {Batterman, Robert W},
	Journal = {No{\^u}s},
	Number = {4},
	Pages = {858--873},
	Publisher = {Wiley Online Library},
	Title = {Autonomy of theories: An explanatory problem},
	Volume = {52},
	Year = {2018}}

@article{Dawid:2021,
	Author = {Dawid, Richard and McCoy, CD},
	Title = {Testability and Viability: Is Inflationary Cosmology ``Scientific"?},
        url={http://philsciarchive.pitt.edu/19335/},
	Year = {2021}}

@article{Mccoy:2015,
	Author = {McCoy, Casey D},
	Journal = {Studies in History and Philosophy of Science Part B: Studies in History and Philosophy of Modern Physics},
	Pages = {23--36},
	Publisher = {Elsevier},
	Title = {Does inflation solve the hot big bang model's fine-tuning problems?},
	Volume = {51},
	Year = {2015}}

@article{Azhar:2021,
	Author = {Azhar, Feraz and Loeb, Abraham},
	Journal = {Foundations of Physics},
	Number = {5},
	Pages = {1--36},
	Publisher = {Springer},
	Title = {Finely tuned models sacrifice explanatory depth},
	Volume = {51},
	Year = {2021}}

@incollection{azhar:2017,
	Author = {Azhar, Feraz and Butterfield, Jeremy},
	Booktitle = {The Routledge Handbook of Scientific Realism},
	Pages = {304--320},
	Publisher = {Routledge},
	Title = {Scientific realism and primordial cosmology},
	Year = {2017}}

@article{COBE,
author = {Smoot, George F. },
title = {COBE observations and results},
journal = {AIP Conference Proceedings},
volume = {476},
number = {1},
pages = {1-10},
year = {1999},
doi = {10.1063/1.59326}
}

@article{WMAP,
  title={Nine-year Wilkinson Microwave Anisotropy Probe (WMAP) observations: final maps and results},
  author={Bennett, Charles L and Larson, Davin and Weiland, Janet L and Jarosik, N and Hinshaw, G and Odegard, N and Smith, KM and Hill, RS and Gold, B and Halpern, M and others},
  journal={The Astrophysical Journal Supplement Series},
  volume={208},
  number={2},
  pages={20},
  year={2013},
  publisher={IOP Publishing}
}

@article{Planck,
  title={Planck 2018 results-VI. Cosmological parameters},
  author={Aghanim, Nabila and Akrami, Yashar and Ashdown, Mark and Aumont, J and Baccigalupi, C and Ballardini, M and Banday, AJ and Barreiro, RB and Bartolo, N and Basak, S and others},
  journal={Astronomy \& Astrophysics},
  volume={641},
  pages={A6},
  year={2020},
  publisher={EDP sciences}
}

@article{Perlmutter,
  title={Measurements* of the Cosmological Parameters $\Omega$ and $\Lambda$ from the First Seven Supernovae at z  0.35},
  author={Perlmutter, Saul and Gabi, S and Goldhaber, G and Goobar, A and Groom, DE and Hook, IM and Kim, AG and Kim, MY and Lee, JC and Pain, R and others},
  journal={The astrophysical journal},
  volume={483},
  number={2},
  pages={565},
  year={1997},
  publisher={IOP Publishing}
}

@article{aubourg2015cosmological,
  title={Cosmological implications of baryon acoustic oscillation measurements},
  author={Aubourg, {\'E}ric and Bailey, Stephen and Bautista, Julian E and Beutler, Florian and Bhardwaj, Vaishali and Bizyaev, Dmitry and Blanton, Michael and Blomqvist, Michael and Bolton, Adam S and Bovy, Jo and others},
  journal={Physical Review D},
  volume={92},
  number={12},
  pages={123516},
  year={2015},
  publisher={APS}
}

@article{Weinbergforest,
author = {Weinberg, David H. and Davé, Romeel and Katz, Neal and Kollmeier, Juna A. },
title = {The Lyman‐$\alpha$ Forest as a Cosmological Tool},
journal = {AIP Conference Proceedings},
volume = {666},
number = {1},
pages = {157-169},
year = {2003},
doi = {10.1063/1.1581786}
}

@article{Ellislensing,
author = {Ellis, Richard},
year = {2010},
month = {03},
pages = {967-87},
title = {Gravitational lensing: A unique probe of dark matter and dark energy},
volume = {368},
journal = {Philosophical transactions. Series A, Mathematical, physical, and engineering sciences},
doi = {10.1098/rsta.2009.0209}
}

@article{Rubinreview,
author = {Sofue, Yoshiaki and Rubin, Vera},
title = {Rotation Curves of Spiral Galaxies},
journal = {Annual Review of Astronomy and Astrophysics},
volume = {39},
number = {1},
pages = {137-174},
year = {2001},
doi = {10.1146/annurev.astro.39.1.137}
}

@article{Allenreview,
author = {Allen, Steven W. and Evrard, August E. and Mantz, Adam B.},
title = {Cosmological Parameters from Observations of Galaxy Clusters},
journal = {Annual Review of Astronomy and Astrophysics},
volume = {49},
number = {1},
pages = {409-470},
year = {2011},
doi = {10.1146/annurev-astro-081710-102514}
}

@article{martin2014encyclopaedia,
  title={Encyclop{\ae}dia inflationaris},
  author={Martin, Jerome and Ringeval, Christophe and Vennin, Vincent},
  journal={Physics of the Dark Universe},
  volume={5},
  pages={75--235},
  year={2014},
  publisher={Elsevier}
}

@article{ijjas2018bouncing,
  title={Bouncing Cosmology made simple},
  author={Ijjas, Anna and Steinhardt, Paul J},
  journal={Classical and Quantum Gravity},
  volume={35},
  number={13},
  pages={135004},
  year={2018},
  publisher={IOP Publishing}
}

@inproceedings{baumann2009tasi,
    author = "Baumann, Daniel",
    title = "{Inflation}",
    booktitle = "{Theoretical Advanced Study Institute in Elementary Particle Physics}: {Physics of the Large and the Small}",
    eprint = "0907.5424",
    archivePrefix = "arXiv",
    primaryClass = "hep-th",
    reportNumber = "TASI-2009",
    doi = "10.1142/9789814327183_0010",
    pages = "523--686",
    year = "2011"
}

@article{andrei2022end,
    author = "Andrei, Cosmin and Ijjas, Anna and Steinhardt, Paul J.",
    title = "{Rapidly descending dark energy and the end of cosmic expansion}",
    eprint = "2201.07704",
    archivePrefix = "arXiv",
    primaryClass = "astro-ph.CO",
    doi = "10.1073/pnas.2200539119",
    journal = "Proc. Nat. Acad. Sci.",
    volume = "119",
    number = "15",
    pages = "e2200539119",
    year = "2022"
}

@article{steinhardt2002cosmic,
  title={Cosmic evolution in a cyclic universe},
  author={Steinhardt, Paul J and Turok, Neil},
  journal={Physical Review D},
  volume={65},
  number={12},
  pages={126003},
  year={2002},
  publisher={APS}
}

@article{steinhardt2002cyclic,
  title={A cyclic model of the universe},
  author={Steinhardt, Paul J and Turok, Neil},
  journal={Science},
  volume={296},
  number={5572},
  pages={1436--1439},
  year={2002},
  publisher={American Association for the Advancement of Science}
}

@book{Maudlin2007-MAUTMW,
	title = {The Metaphysics Within Physics},
	year = {2007},
	publisher = {Oxford University Press},
	author = {Tim Maudlin}
}

@article{SteinhardtGuth,
 ISSN = {00368733, 19467087},
 URL = {http://www.jstor.org/stable/24969372},
 author = {Alan H. Guth and Paul J. Steinhardt},
 journal = {Scientific American},
 number = {5},
 pages = {116--129},
 publisher = {Scientific American, a division of Nature America, Inc.},
 title = {The Inflationary Universe},
 urldate = {2022-06-22},
 volume = {250},
 year = {1984}
}

@article{Schiffrin:2012,
  title={Measure and probability in cosmology},
  author={Schiffrin, Joshua S and Wald, Robert M},
  journal={Physical Review D},
  volume={86},
  number={2},
  pages={023521},
  year={2012},
  publisher={APS}
}

@article{ijjas2020robustness,
  title={Robustness of slow contraction to cosmic initial conditions},
  author={Ijjas, Anna and Cook, William G and Pretorius, Frans and Steinhardt, Paul J and Davies, Elliot Y},
  journal={Journal of Cosmology and Astroparticle Physics},
  volume={2020},
  number={08},
  pages={030},
  year={2020},
  publisher={IOP Publishing}
}

@article{Harrison,
  title = {Fluctuations at the Threshold of Classical Cosmology},
  author = {Harrison, E. R.},
  journal = {Phys. Rev. D},
  volume = {1},
  issue = {10},
  pages = {2726--2730},
  numpages = {0},
  year = {1970},
  %month = {May},
  publisher = {American Physical Society},
  doi = {10.1103/PhysRevD.1.2726},
  url = {https://link.aps.org/doi/10.1103/PhysRevD.1.2726}
}

@ARTICLE{PeeblesYu,
       author = {{Peebles}, P.~J.~E. and {Yu}, J.~T.},
        title = "{Primeval Adiabatic Perturbation in an Expanding Universe}",
      journal = {apj},
         year = 1970,
        month = dec,
       volume = {162},
        pages = {815},
          doi = {10.1086/150713},
       adsurl = {https://ui.adsabs.harvard.edu/abs/1970ApJ...162..815P}
}

@ARTICLE{Zeldovich,
       author = {{Zeldovich}, Yaa B.},
        title = "{A hypothesis, unifying the structure and the entropy of the Universe}",
      journal = {mnras},
         year = 1972,
        month = jan,
       volume = {160},
        pages = {1P},
          doi = {10.1093/mnras/160.1.1P},
       adsurl = {https://ui.adsabs.harvard.edu/abs/1972MNRAS.160P...1Z}
}

@article{Guth1982,
  title = {Fluctuations in the New Inflationary Universe},
  author = {Guth, Alan H. and Pi, So-Young},
  journal = {Phys. Rev. Lett.},
  volume = {49},
  issue = {15},
  pages = {1110--1113},
  numpages = {0},
  year = {1982},
  %month = {Oct},
  publisher = {American Physical Society},
  doi = {10.1103/PhysRevLett.49.1110},
  url = {https://link.aps.org/doi/10.1103/PhysRevLett.49.1110}
}

@article{HAWKING1982295,
title = {The development of irregularities in a single bubble inflationary universe},
journal = {Physics Letters B},
volume = {115},
number = {4},
pages = {295-297},
year = {1982},
issn = {0370-2693},
doi = {https://doi.org/10.1016/0370-2693(82)90373-2},
url = {https://www.sciencedirect.com/science/article/pii/0370269382903732},
author = {S.W. Hawking}
}

@article{Bardeen1983,
  title = {Spontaneous creation of almost scale-free density perturbations in an inflationary universe},
  author = {Bardeen, James M. and Steinhardt, Paul J. and Turner, Michael S.},
  journal = {Phys. Rev. D},
  volume = {28},
  issue = {4},
  pages = {679--693},
  numpages = {0},
  year = {1983},
  %month = {Aug},
  publisher = {American Physical Society},
  doi = {10.1103/PhysRevD.28.679}
}

@article{Zaldarriaga:1996xe,
    author = "Zaldarriaga, Matias and Seljak, Uros",
    title = "{An all sky analysis of polarization in the microwave background}",
    eprint = "astro-ph/9609170",
    archivePrefix = "arXiv",
    doi = "10.1103/PhysRevD.55.1830",
    journal = "Phys. Rev. D",
    volume = "55",
    pages = "1830--1840",
    year = "1997"
}

@article{Baumann:2009mq,
    author = "Baumann, Daniel and Zaldarriaga, Matias",
    title = "{Causality and Primordial Tensor Modes}",
    eprint = "0901.0958",
    archivePrefix = "arXiv",
    primaryClass = "astro-ph.CO",
    doi = "10.1088/1475-7516/2009/06/013",
    journal = "JCAP",
    volume = "06",
    pages = "013",
    year = "2009"
}

@article{Ijjas:2019pyf,
    author = "Ijjas, Anna and Steinhardt, Paul J.",
    title = "{A new kind of cyclic universe}",
    eprint = "1904.08022",
    archivePrefix = "arXiv",
    primaryClass = "gr-qc",
    doi = "10.1016/j.physletb.2019.06.056",
    journal = "Phys. Lett. B",
    volume = "795",
    pages = "666--672",
    year = "2019"
}

@article{Chowdhury:2019otk,
    author = "Chowdhury, Debika and Martin, J\'er\^ome and Ringeval, Christophe and Vennin, Vincent",
    title = "{Assessing the scientific status of inflation after Planck}",
    eprint = "1902.03951",
    archivePrefix = "arXiv",
    primaryClass = "astro-ph.CO",
    doi = "10.1103/PhysRevD.100.083537",
    journal = "Phys. Rev. D",
    volume = "100",
    number = "8",
    pages = "083537",
    year = "2019"
}

@article{Martin:2015dha,
    author = "Martin, Jerome",
    editor = "Fabris, J\'ulio C. and Piattella, Oliver F. and Rodrigues, Davi C. and Velten, Hermano E. S. and Zimdahl, Winfried",
    title = "{The Observational Status of Cosmic Inflation after Planck}",
    eprint = "1502.05733",
    archivePrefix = "arXiv",
    primaryClass = "astro-ph.CO",
    doi = "10.1007/978-3-319-44769-8_2",
    journal = "Astrophys. Space Sci. Proc.",
    volume = "45",
    pages = "41--134",
    year = "2016"
}

@article{Planck:2018jri,
    author = "Akrami, Y. and others",
    collaboration = "Planck",
    title = "{Planck 2018 results. X. Constraints on inflation}",
    eprint = "1807.06211",
    archivePrefix = "arXiv",
    primaryClass = "astro-ph.CO",
    doi = "10.1051/0004-6361/201833887",
    journal = "Astron. Astrophys.",
    volume = "641",
    pages = "A10",
    year = "2020"
}

@article{Wolf:2019hzy,
    author = "Wolf, William J. and Lagos, Macarena",
    title = "{Cosmological Instabilities and the Role of Matter Interactions in Dynamical Dark Energy Models}",
    eprint = "1908.03212",
    archivePrefix = "arXiv",
    primaryClass = "gr-qc",
    doi = "10.1103/PhysRevD.100.084035",
    journal = "Phys. Rev. D",
    volume = "100",
    number = "8",
    pages = "084035",
    year = "2019"
}

@article{Lehners:2007ac,
    author = "Lehners, Jean-Luc and McFadden, Paul and Turok, Neil and Steinhardt, Paul J.",
    title = "{Generating ekpyrotic curvature perturbations before the big bang}",
    eprint = "hep-th/0702153",
    archivePrefix = "arXiv",
    reportNumber = "DAMTP-2007-14, ITFA-2007-12",
    doi = "10.1103/PhysRevD.76.103501",
    journal = "Phys. Rev. D",
    volume = "76",
    pages = "103501",
    year = "2007"
}

@article{Creminelli:2004jg,
    author = "Creminelli, Paolo and Nicolis, Alberto and Zaldarriaga, Matias",
    title = "{Perturbations in bouncing cosmologies: Dynamical attractor versus scale invariance}",
    eprint = "hep-th/0411270",
    archivePrefix = "arXiv",
    reportNumber = "HUTP-04-A043",
    doi = "10.1103/PhysRevD.71.063505",
    journal = "Phys. Rev. D",
    volume = "71",
    pages = "063505",
    year = "2005"
}

@article{Levy:2015awa,
    author = "Levy, Aaron M. and Ijjas, Anna and Steinhardt, Paul J.",
    title = "{Scale-invariant perturbations in ekpyrotic cosmologies without fine-tuning of initial conditions}",
    eprint = "1506.01011",
    archivePrefix = "arXiv",
    primaryClass = "astro-ph.CO",
    doi = "10.1103/PhysRevD.92.063524",
    journal = "Phys. Rev. D",
    volume = "92",
    number = "6",
    pages = "063524",
    year = "2015"
}

@article{Brandenberger:2016vhg,
    author = "Brandenberger, Robert and Peter, Patrick",
    title = "{Bouncing Cosmologies: Progress and Problems}",
    eprint = "1603.05834",
    archivePrefix = "arXiv",
    primaryClass = "hep-th",
    doi = "10.1007/s10701-016-0057-0",
    journal = "Found. Phys.",
    volume = "47",
    number = "6",
    pages = "797--850",
    year = "2017"
}

@article{Ijjas:2016tpn,
    author = "Ijjas, Anna and Steinhardt, Paul J.",
    title = "{Classically stable nonsingular cosmological bounces}",
    eprint = "1606.08880",
    archivePrefix = "arXiv",
    primaryClass = "gr-qc",
    doi = "10.1103/PhysRevLett.117.121304",
    journal = "Phys. Rev. Lett.",
    volume = "117",
    number = "12",
    pages = "121304",
    year = "2016"
}

@article{Ijjas:2016vtq,
    author = "Ijjas, Anna and Steinhardt, Paul J.",
    title = "{Fully stable cosmological solutions with a non-singular classical bounce}",
    eprint = "1609.01253",
    archivePrefix = "arXiv",
    primaryClass = "gr-qc",
    doi = "10.1016/j.physletb.2016.11.047",
    journal = "Phys. Lett. B",
    volume = "764",
    pages = "289--294",
    year = "2017"
}

@article{Horndeski:1974wa,
    author = "Horndeski, Gregory Walter",
    title = "{Second-order scalar-tensor field equations in a four-dimensional space}",
    doi = "10.1007/BF01807638",
    journal = "Int. J. Theor. Phys.",
    volume = "10",
    pages = "363--384",
    year = "1974"
}

@article{Martin:2003kp,
    author = "Martin, Jerome and Brandenberger, Robert",
    title = "{On the dependence of the spectra of fluctuations in inflationary cosmology on transPlanckian physics}",
    eprint = "hep-th/0305161",
    archivePrefix = "arXiv",
    doi = "10.1103/PhysRevD.68.063513",
    journal = "Phys. Rev. D",
    volume = "68",
    pages = "063513",
    year = "2003"
}

@article{Brandenberger:2012aj,
    author = "Brandenberger, Robert H. and Martin, Jerome",
    title = "{Trans-Planckian Issues for Inflationary Cosmology}",
    eprint = "1211.6753",
    archivePrefix = "arXiv",
    primaryClass = "astro-ph.CO",
    doi = "10.1088/0264-9381/30/11/113001",
    journal = "Class. Quant. Grav.",
    volume = "30",
    pages = "113001",
    year = "2013"
}

@article{Cai:2014bea,
    author = "Cai, Yi-Fu",
    title = "{Exploring Bouncing Cosmologies with Cosmological Surveys}",
    eprint = "1405.1369",
    archivePrefix = "arXiv",
    primaryClass = "hep-th",
    doi = "10.1007/s11433-014-5512-3",
    journal = "Sci. China Phys. Mech. Astron.",
    volume = "57",
    pages = "1414--1430",
    year = "2014"
}

@article{Brandenberger:2021pzy,
    author = "Brandenberger, Robert",
    title = "{Trans-Planckian Censorship Conjecture and Early Universe Cosmology}",
    eprint = "2102.09641",
    archivePrefix = "arXiv",
    primaryClass = "hep-th",
    doi = "10.31526/lhep.2021.198",
    journal = "LHEP",
    volume = "2021",
    pages = "198",
    year = "2021"
}

@article{Bedroya:2019tba,
    author = "Bedroya, Alek and Brandenberger, Robert and Loverde, Marilena and Vafa, Cumrun",
    title = "{Trans-Planckian Censorship and Inflationary Cosmology}",
    eprint = "1909.11106",
    archivePrefix = "arXiv",
    primaryClass = "hep-th",
    doi = "10.1103/PhysRevD.101.103502",
    journal = "Phys. Rev. D",
    volume = "101",
    number = "10",
    pages = "103502",
    year = "2020"
}

@article{Bedroya:2019snp,
    author = "Bedroya, Alek and Vafa, Cumrun",
    title = "{Trans-Planckian Censorship and the Swampland}",
    eprint = "1909.11063",
    archivePrefix = "arXiv",
    primaryClass = "hep-th",
    doi = "10.1007/JHEP09(2020)123",
    journal = "JHEP",
    volume = "09",
    pages = "123",
    year = "2020"
}

@article{Ijjas:2013vea,
    author = "Ijjas, Anna and Steinhardt, Paul J. and Loeb, Abraham",
    title = "{Inflationary paradigm in trouble after Planck2013}",
    eprint = "1304.2785",
    archivePrefix = "arXiv",
    primaryClass = "astro-ph.CO",
    doi = "10.1016/j.physletb.2013.05.023",
    journal = "Phys. Lett. B",
    volume = "723",
    pages = "261--266",
    year = "2013"
}

@article{Rubakov:2014jja,
    author = "Rubakov, V. A.",
    title = "{The Null Energy Condition and its violation}",
    eprint = "1401.4024",
    archivePrefix = "arXiv",
    primaryClass = "hep-th",
    reportNumber = "INR-TH-2014-1",
    doi = "10.3367/UFNe.0184.201402b.0137",
    journal = "Phys. Usp.",
    volume = "57",
    pages = "128--142",
    year = "2014"
}

@article{Battefeld:2014uga,
    author = "Battefeld, D. and Peter, Patrick",
    title = "{A Critical Review of Classical Bouncing Cosmologies}",
    eprint = "1406.2790",
    archivePrefix = "arXiv",
    primaryClass = "astro-ph.CO",
    doi = "10.1016/j.physrep.2014.12.004",
    journal = "Phys. Rept.",
    volume = "571",
    pages = "1--66",
    year = "2015"
}

@article{Borde:2001nh,
    author = "Borde, Arvind and Guth, Alan H. and Vilenkin, Alexander",
    title = "{Inflationary space-times are incompletein past directions}",
    eprint = "gr-qc/0110012",
    archivePrefix = "arXiv",
    reportNumber = "MIT-CTP-3183",
    doi = "10.1103/PhysRevLett.90.151301",
    journal = "Phys. Rev. Lett.",
    volume = "90",
    pages = "151301",
    year = "2003"
}

@article{Ijjas:2014nta,
    author = "Ijjas, Anna and Steinhardt, Paul J. and Loeb, Abraham",
    title = "{Inflationary schism}",
    eprint = "1402.6980",
    archivePrefix = "arXiv",
    primaryClass = "astro-ph.CO",
    doi = "10.1016/j.physletb.2014.07.012",
    journal = "Phys. Lett. B",
    volume = "736",
    pages = "142--146",
    year = "2014"
}

@article{Cai:2012va,
    author = "Cai, Yi-Fu and Easson, Damien A. and Brandenberger, Robert",
    title = "{Towards a Nonsingular Bouncing Cosmology}",
    eprint = "1206.2382",
    archivePrefix = "arXiv",
    primaryClass = "hep-th",
    doi = "10.1088/1475-7516/2012/08/020",
    journal = "JCAP",
    volume = "08",
    pages = "020",
    year = "2012"
}

@article{Easson:2011zy,
    author = "Easson, Damien A. and Sawicki, Ignacy and Vikman, Alexander",
    title = "{G-Bounce}",
    eprint = "1109.1047",
    archivePrefix = "arXiv",
    primaryClass = "hep-th",
    reportNumber = "CERN-PH-TH-2011-203",
    doi = "10.1088/1475-7516/2011/11/021",
    journal = "JCAP",
    volume = "11",
    pages = "021",
    year = "2011"
}

@article{Libanov:2016kfc,
    author = "Libanov, M. and Mironov, S. and Rubakov, V.",
    title = "{Generalized Galileons: instabilities of bouncing and Genesis cosmologies and modified Genesis}",
    eprint = "1605.05992",
    archivePrefix = "arXiv",
    primaryClass = "hep-th",
    reportNumber = "INR-TH-2016-014",
    doi = "10.1088/1475-7516/2016/08/037",
    journal = "JCAP",
    volume = "08",
    pages = "037",
    year = "2016"
}

@article{Kobayashi:2016xpl,
    author = "Kobayashi, Tsutomu",
    title = "{Generic instabilities of nonsingular cosmologies in Horndeski theory: A no-go theorem}",
    eprint = "1606.05831",
    archivePrefix = "arXiv",
    primaryClass = "hep-th",
    reportNumber = "RUP-16-19",
    doi = "10.1103/PhysRevD.94.043511",
    journal = "Phys. Rev. D",
    volume = "94",
    number = "4",
    pages = "043511",
    year = "2016"
}

@article{Cai:2016thi,
    author = "Cai, Yong and Wan, Youping and Li, Hai-Guang and Qiu, Taotao and Piao, Yun-Song",
    title = "{The Effective Field Theory of nonsingular cosmology}",
    eprint = "1610.03400",
    archivePrefix = "arXiv",
    primaryClass = "gr-qc",
    doi = "10.1007/JHEP01(2017)090",
    journal = "JHEP",
    volume = "01",
    pages = "090",
    year = "2017"
}

@article{Cai:2017tku,
    author = "Cai, Yong and Li, Hai-Guang and Qiu, Taotao and Piao, Yun-Song",
    title = "{The Effective Field Theory of nonsingular cosmology: II}",
    eprint = "1701.04330",
    archivePrefix = "arXiv",
    primaryClass = "gr-qc",
    doi = "10.1140/epjc/s10052-017-4938-y",
    journal = "Eur. Phys. J. C",
    volume = "77",
    number = "6",
    pages = "369",
    year = "2017"
}

@article{Cai:2017dyi,
    author = "Cai, Yong and Piao, Yun-Song",
    title = "{A covariant Lagrangian for stable nonsingular bounce}",
    eprint = "1705.03401",
    archivePrefix = "arXiv",
    primaryClass = "gr-qc",
    doi = "10.1007/JHEP09(2017)027",
    journal = "JHEP",
    volume = "09",
    pages = "027",
    year = "2017"
}

@article{Creminelli:2016zwa,
    author = "Creminelli, Paolo and Pirtskhalava, David and Santoni, Luca and Trincherini, Enrico",
    title = "{Stability of Geodesically Complete Cosmologies}",
    eprint = "1610.04207",
    archivePrefix = "arXiv",
    primaryClass = "hep-th",
    doi = "10.1088/1475-7516/2016/11/047",
    journal = "JCAP",
    volume = "11",
    pages = "047",
    year = "2016"
}

@article{Linde:1983gd,
    author = "Linde, Andrei D.",
    title = "{Chaotic Inflation}",
    doi = "10.1016/0370-2693(83)90837-7",
    journal = "Phys. Lett. B",
    volume = "129",
    pages = "177--181",
    year = "1983"
}

\end{document}